\documentclass[paper]{JHEP3}
\usepackage{booktabs}
\usepackage{psfig}
\usepackage{epsfig,bm,amsmath}
\usepackage{enumerate}

\preprint{Cavendish--HEP--06/27\\
KA--TP--10--2006}
{\title{A Positive-Weight Next-to-Leading-Order Monte Carlo for \boldmath{$e^+e^-$} Annihilation to
  Hadrons}}
\author{Oluseyi Latunde-Dada$^1$, Stefan Gieseke$^2$, Bryan Webber$^3$\\
  $^{1,3}$Cavendish Laboratory, University of Cambridge,\\
  JJ Thomson Avenue, Cambridge CB3 0HE, U.K.\\
  $^2$Institut f\"{u}r Theoretische Physik, Universit\"{a}t Karlsruhe, 76128 Karlsruhe, Germany.\\
  $^1$E-mail: \email{seyi@hep.phy.cam.ac.uk}\\
  $^2$E-mail: \email{gieseke@particle.uni-karlsruhe.de}\\
  $^3$E-mail: \email{webber@hep.phy.cam.ac.uk}}

\abstract{We apply the positive-weight Monte Carlo method of Nason for simulating QCD processes accurate to
  Next-To-Leading Order to the case of $e^+e^-$ annihilation to hadrons. The
  method entails the generation of the hardest gluon emission first and then subsequently
  adding a `truncated' shower before the emission. We have interfaced our result to the
  {\tt Herwig++} shower Monte Carlo program and obtained better results than those obtained
  with {\tt Herwig++} at leading order with a matrix element correction.
}


\begin{document}


\section{Introduction}
\label{sec:introduction}
Matching next-to-leading order (NLO) calculations to shower Monte Carlo (SMC) models is
highly non-trivial. One method (MC@NLO) has been proposed in \cite{Frixione:2002ik} and has been
successfully applied to several processes \cite{Frixione:2003ei, Frixione:2006gn} in connection with the {\tt FORTRAN
 HERWIG} parton shower. One drawback of the method is the generation of negative weighted
events which are unphysical.

An alternative method proposed in \cite{Nason:2004rx} overcomes the negative weight problem. It has
successfully been applied to Z pair hadroproduction \cite{Nason:2006hf} and involves the generation of
the hardest radiation independently of the SMC model used to generate the parton
shower. To preserve the soft radiation distribution, the addition of a `truncated shower'
of soft coherent radiation before the hardest emission
is necessary. 
In this report, we will be implementing this method for electron-positron
annihilation to hadrons at a centre-of-mass energy of $91.2$ GeV. The SMC we will be employing is
{\tt Herwig++} \cite{Gieseke:2003hm} which will
perform the rest of the showering, hadronization and decays.

In Section \ref{sec:hardest-emission}, we discuss the generation of the hardest emission
according to the matrix element via a modified Sudakov form factor. In Section \ref{sec:trunc}, we go on to describe the
generation of a simplified form of the `truncated' shower. We determine the probability for
the emission of at most one gluon before the hardest emission and re-shuffle the
momenta of the outgoing particles accordingly. This is a reasonable first approximation
since the probability of emitting an extra gluon is small.

In Section \ref{sec:results}, we present our results compared with the measured data and
those obtained using  {\tt Herwig++} with a matrix element correction. Finally in
Section \ref{sec:conclusions}, we
summarize our conclusions. 
\section{Hardest emission generation}
\label{sec:hardest-emission}
\subsection{Hardest emission cross section}
\label{sec:cross-section}
The order-$\alpha_{s}$ differential cross section for the process $e^+e^- \rightarrow
q\bar{q}g$, neglecting the quark masses, can be written as 
\begin{equation}
\centering
\label{eq:W}
R(x,y)=\sigma_{0}W(x,y)=\sigma_{0}\frac{2{\alpha}_s}{3{\pi}}\frac{x^2+y^2}{(1-x)(1-y)}
\end{equation}
where $\sigma_{0}$ is the Born cross section, $W=R/\sigma_{0}$, $x$ and $y$ are the energy fractions of the quark and antiquark and $2-x-y$ is the energy fraction of the gluon.
$R$ has collinear singularities when the quark or antiquark is aligned with the
gluon. When combined with the virtual corrections, these singularities can be integrated to give the well known total cross section to order-$\alpha_{s}$,
\begin{equation}
\centering
\label{eq:NLO}
\sigma_{NLO}=\sigma_{0}\left[1+\frac{\alpha_{s}}{\pi}\right]\;.
\end{equation}
From \cite{Nason:2004rx}, we can write the cross section for the hardest gluon emission event as
\begin{equation}
\centering
\label{eq:sig}
d \sigma=\sum \bar {B}(v) d \Phi_{v}\left[\Delta^{(NLO)}_{R}(0)+\Delta^{(NLO)}_{R}(p_T)\frac{R(v,r)}{B(v)}d \Phi_{r}\right]
\end{equation}
where $B(v)$ is the Born cross section and $v$ is the Born variable, which in this case is the
angle $\theta$ between the $e^+e^-$ beam axis and the $q\bar{q}$ axis. $r$ represents the
radiation variables ($x$ and $y$ for our specific case), $d \Phi_{v}$ and $d
\Phi_{r}$ are the Born and real emission phase spaces respectively. 

$\Delta_{R}^{NLO}(p_T)$ is the modified Sudakov form factor for the hardest emission with
transverse momentum $p_T$, as indicated by the Heaviside function in the exponent of
(\ref{eq:dnlo}),
\begin{equation}
\centering
\label{eq:dnlo}
\Delta_{R}^{NLO}(p_T)=\exp \left[-\int d\Phi{r}\frac{R(v,r)}{B(v)}\Theta(k_T(v,r)-p_T)\right]\;.
\end{equation}
Furthermore, 
\begin{equation}
\centering
\label{eq:B}
\bar{B}(v)=B(v)+V(v)+\int (R(v,r)-C(v,r))d \Phi_{r}\;.
\end{equation}
$\bar{B}(v)$ is the sum of the Born, $B(v)$, virtual, $V(v)$ and real, $R(v,r)$ terms, (with some
counter-terms, $C(v,r)$). It overcomes the problem of negative weights
since in the region where $\bar{B}(v)$ is negative, the NLO negative terms must have overcome the Born
term and hence perturbation theory must have failed.
Now explicitly for $e^+e^-$ annihilation,
\begin{equation}
\centering
\label{eq:delta}
\Delta_{R}^{NLO}(p_T)=\exp \left[-\int dx\,dy \frac{2{\alpha}_s}{3{\pi}}\frac{x^2+y^2}{(1-x)(1-y)}\Theta(k_T(x,y))-p_T)\right]
\end{equation}
where $x$ and $y$ are the energy fractions of the quark and antiquark and we define
\begin{equation}
\centering
\label{eq:kT}
k_T(x,y)=\sqrt{s\frac{(1-x)(1-y)(x+y-1)}{\max(x,y)^2}}
\end{equation}
and $s$ as the square of the center-of-mass energy. $k_T$ is the transverse momentum of
the hardest emitted gluon relative to the splitting axis, as illustrated in Figure \ref{fig:kt}
below.
\begin{figure}[!ht]
\begin{center}
\psfig{figure=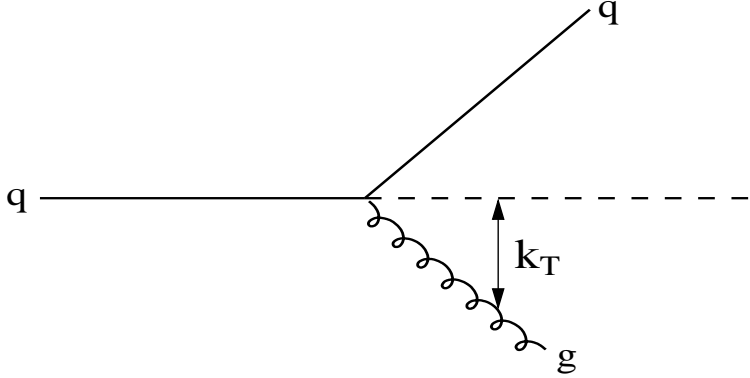,%
width=4in,height=2in,angle=0}
\end{center}
\caption{Transverse momentum, $k_T$.}
\label{fig:kt}
\end{figure}

\subsection{Generation of radiation variables, $x$ and $y$}
\label{sec:xy}

\begin{figure}[!ht]
\begin{center}
\psfig{figure=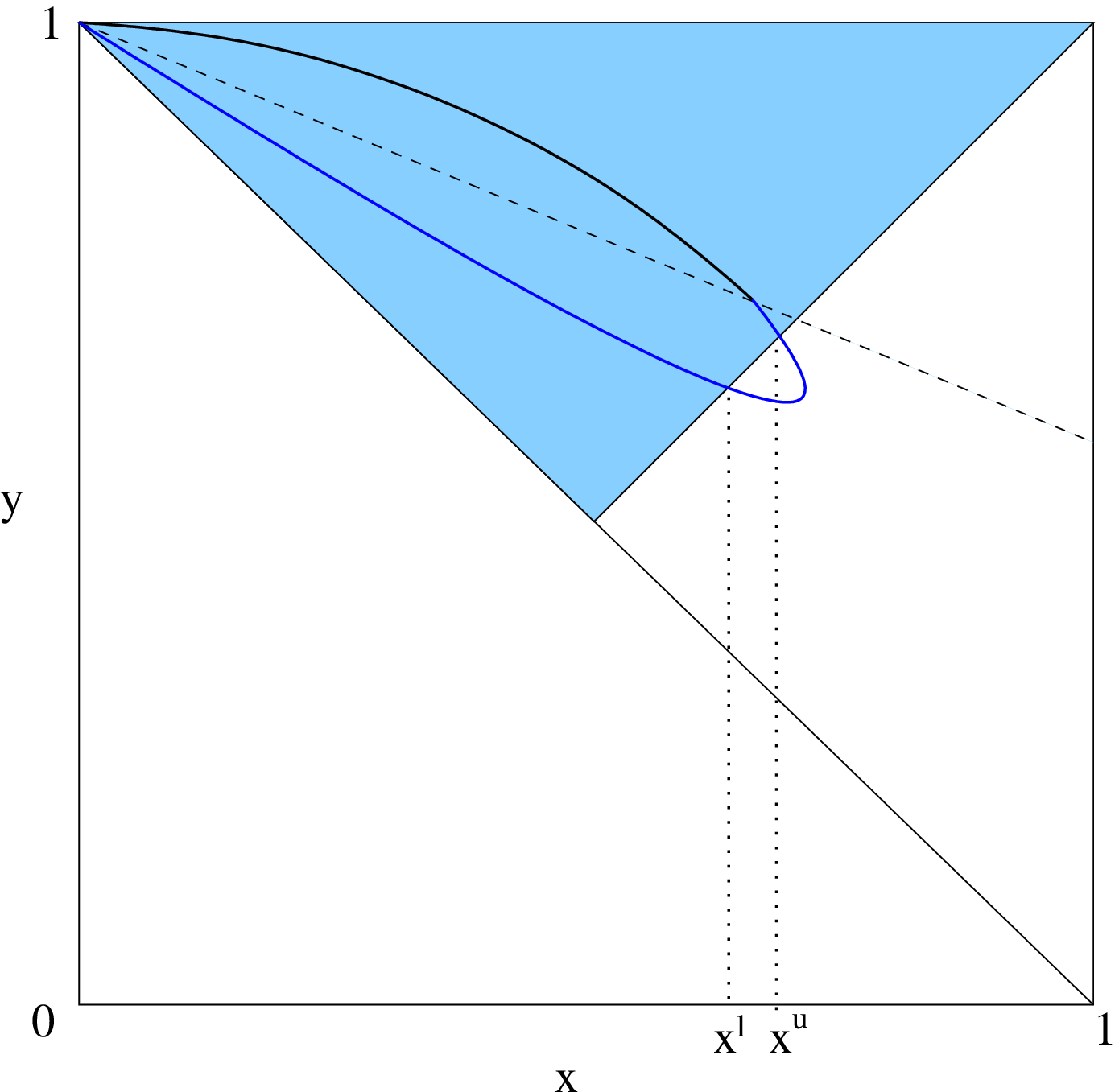,%
width=3in,height=3in,angle=0}
\end{center}
\caption{Phase space and $y$ solutions for $\kappa>0.08333$ in the region $x>y$.}
\label{fig:gt}
\end{figure}

From (\ref{eq:W}) and (\ref{eq:sig}), it can be deduced that the radiation variables are
to be generated according to the
probability distribution
\begin{equation}
\centering
\label{eq:delW}
\Delta^{W}(k_T)W(x,y)dx\,dy
\end{equation}
where in the particular case of $e^+e^-$ annihilation, $\Delta^{W}(k_T)$ and $W(x,y)$ are given
in (\ref{eq:delta}) and (\ref{eq:W}) respectively. However for ease
of integration, we will use a function 
\begin{equation}
\centering
\label{eq:U}
U(x,y)=\frac{2{\alpha}_s}{3{\pi}}\frac{2}{(1-x)(1-y)}\ge W(x,y)
\end{equation}
in place of $W(x,y)$. As we shall see later, the true distribution is recovered using the veto technique.
The variables $x$ and $y$ are then generated according to 
\begin{equation}
\centering
\label{eq:delU}
\Delta^{U}(k_T)U(x,y)=U(x,y)\exp \left[-\int U(x,y){\Theta}(k_T(x,y)-p_T)dx \,dy\right]\;.
\end{equation}
This is outlined in the following steps:
\begin{enumerate}[i)]
\item
Set $p_{\rm max}={k_{T}}_{\rm max}$.\\
\item
For a random number, $n$ between $0$ and $1$, solve the equation below for $p_T$ 
\begin{equation}
\centering
\label{eq:n}
n=\frac{{\Delta^{U}(p_T)}}{\Delta^{U}(p_{\rm max})}\;.
\end{equation}
\item
Generate the variables $x$ and $y$ according to the distribution
\begin{equation}
\centering
\label{eq:Udel}
U(x,y){\delta}(k_T(x,y)-p_T)\;.
\end{equation}
\item
Accept the generated value of $p_T$ with probability $W/U$. If the event is rejected set
$p_{\rm max}=p_T$ and go to step ii).
\end{enumerate}
Now for  $e^+e^-$ annihilation,
\begin{equation}
\centering
\label{eq:intU}
\int U(x,y) {\Theta}(k_T(x,y)-p_T)dx \,dy= \int^{1}_{0} dx \int^{1}_{1-x} dy U(x,y){\Theta}(k_T(x,y)-p_T)\;. 
\end{equation}
In the region where $x>y$, let's define the dimensionless variable, $\kappa$ as
\begin{equation}
\centering
\label{eq:kappa}
\kappa=\frac{k_{T}^{2}}{s}=\frac{(1-x)(1-y)(x+y-1)}{x^2}\;.
\end{equation}
There are 2 solutions for $y$ for each value of $x$ and $\kappa$, i.e $y_1=y$ and
$y_2=2-x-y$.
\begin{equation}
\centering
y_{1,2}=\frac{-2+3x-x^2\mp\sqrt{x^2-2x^3+x^4-4 \kappa x^2+4 \kappa x^3}}{2(x-1)}\;. 
\label{eq:y1y2}
\end{equation}
Exchanging the $y$ variable for $\kappa$ in (\ref{eq:intU}) we find
\begin{eqnarray}
\centering
\label{eq:int}
\int U(x,y) {\Theta}(k_T(x,y)-p_T)dxdy&=& \int^{x_{\rm max}}_{x_{\rm min}} dx
\int^{\kappa_{\rm max}}_{\kappa}
  d \kappa\frac{2{\alpha}_s(\kappa s)}{3{\pi}}\frac{dy}{d \kappa}U(x,\kappa)\nonumber \\
&=&\int^{x_{\rm max}}_{x_{\rm min}} dx \int^{\kappa_{\rm max}}_{\kappa}
  d \kappa\frac{2{\alpha}_s(\kappa
    s)}{3{\pi}}\frac{4}{(1-x+\rho(\kappa,x))\rho(\kappa,x)}\nonumber \\
\end{eqnarray}
where
\begin{equation}
\centering
\label{eq:rho}
\rho(\kappa,x)=\sqrt{(1-x)(1-4\kappa-x)}\;.
\end{equation}
Note that the argument of $\alpha_s$ is the scale ${k_T}^2=\kappa s$ with
$\Lambda_{QCD}=200$ MeV.
Equation (\ref{eq:int}) applies to the region of phase space where $x>y$. For the region $x<y$,
$x$ and $y$ are exchanged in the equation. \\
The $x$ integration can be performed to yield
\begin{eqnarray}
\centering
\label{eq:intg}
\int U(x,y) {\Theta}(k_T(x,y)-p_T)dxdy &=&\int^{\kappa_{\rm max}}_{\kappa} d \kappa
\frac{2{\alpha}_s(\kappa
  s)}{3{\pi}}\Bigg[\frac{\ln(1-x)}{\kappa}+ 
\nonumber \\*&& \phantom{a}
 +\frac{2\ln(\sqrt{1-x}+\sqrt{1-4\kappa-x})}{\kappa}\Bigg]^{x_{\rm max}}_{x_{\rm
   min}}\;.\phantom{aaaa}
\end{eqnarray}
For $\kappa \le 0.08333$, the region of phase space where $x>y$ and the two solutions for
$y$ for
given values of $x$ and $\kappa$  are illustrated in Figure \ref{fig:lt} below. 
\begin{figure}[!ht]
\begin{center}
\psfig{figure=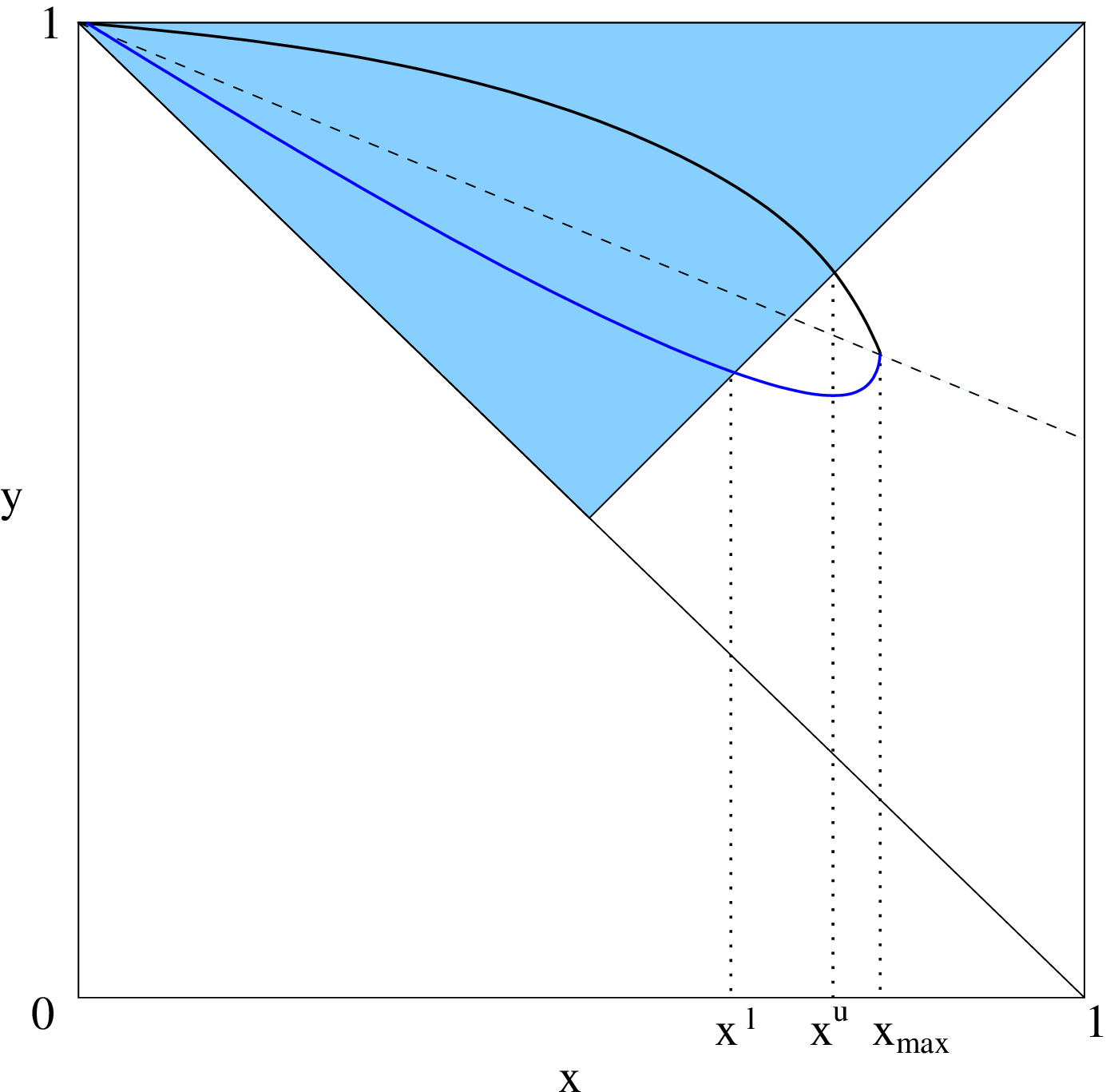,%
width=3in,height=3in,angle=0}
\end{center}
\caption{Phase space and $y$ solutions for $\kappa<0.08333$ in the region $x>y$.}
\label{fig:lt}
\end{figure}
The two solutions lie on either side of the dashed line $y=1-\frac{1}{2}x$ and are
equal when $x=x_{\rm max}=1-4\kappa$ which lies on the dotted line. At $\kappa=0.08333$, the branches
meet along the line $y=x$ and there is only
one solution for $y$ in the region (the lower branch). So for $\kappa>0.08333$, only one $y$
solution exists. In addition there are no $y$ solutions for $\kappa>0.09$. This is
illustrated in Figure \ref{fig:gt}.
Also note that for $x<x^{u}$, there is only one solution for $y$.

In the region where there are two solutions, the integral in (\ref{eq:intg}) is performed along
both branches independently and summed. For the upper branch, $x$ runs from $x^{u}$ to
$x_{\rm max}=1-4\kappa$  while for the lower branch, $x$ runs from $x^{l}$ to $x_{\rm max}$ where
if we define
\clearpage
\begin{eqnarray}
\label{eq:x}
x_a&=&39\kappa-1+\kappa^3+15\kappa^2\,,\nonumber \\
x_b&=&6\sqrt{-33\kappa^2+3\kappa-3\kappa^3}\,,\nonumber \\
x_c&=&\sqrt{{x_a}^2+{x_b}^2}\,,\nonumber \\
x_d&=&\tan^{-1}\left(\frac{x_b}{x_a}\right)\,,\nonumber \\
x_e&=&-\frac{1}{12}x_c^{\frac{1}{3}}\cos\left(\frac{x_d}{3}\right)\,,\nonumber \\
x_f&=&\frac{(-1-\kappa^2-10\kappa)\cos\left(\frac{x_d}{3}\right)}{12x_c^{\frac{1}{3}}}\,,\nonumber \\
x_g&=&\frac{\kappa+5}{6}\,, \nonumber \\
x_h&=&\frac{\sqrt{3}}{12}\sin\left(\frac{x_d}{3}\right)\left(x_c^{\frac{1}{3}}+\frac{1+\kappa^2+10\kappa}{x_c^{\frac{1}{3}}}\right)\,,
\end{eqnarray}
we can write $x^{u}$ and $x^{l}$ as;
\begin{eqnarray}
\centering
\label{eq:xuxl}
x^{u}&=&x_e+x_f+x_g+x_h\,, \nonumber \\
x^{l}&=&x_e+x_f+x_g-x_h \,;
\end{eqnarray}
In the region where there is only one solution for $y$, $x$ runs from $x^{l}$ to $x^{u}$.\\
The $\kappa$ integration can then be performed numerically. Having performed the
integration, values for $\kappa$ and hence $k_T$ are then generated according to steps 1 and 2 in
Section \ref{sec:xy}. In step 3 of Section \ref{sec:xy}, the variables $x$ and $y$ are to be
distributed according to $U(x,y){\delta}(k_T(x,y)-p_T)$. This is the subject of the next section.

\subsection{Distributing $x$ and $y$ according to $U(x,y)$}
\label{sec:Uxy}
\begin{figure}[!ht]
\begin{center}
\psfig{figure=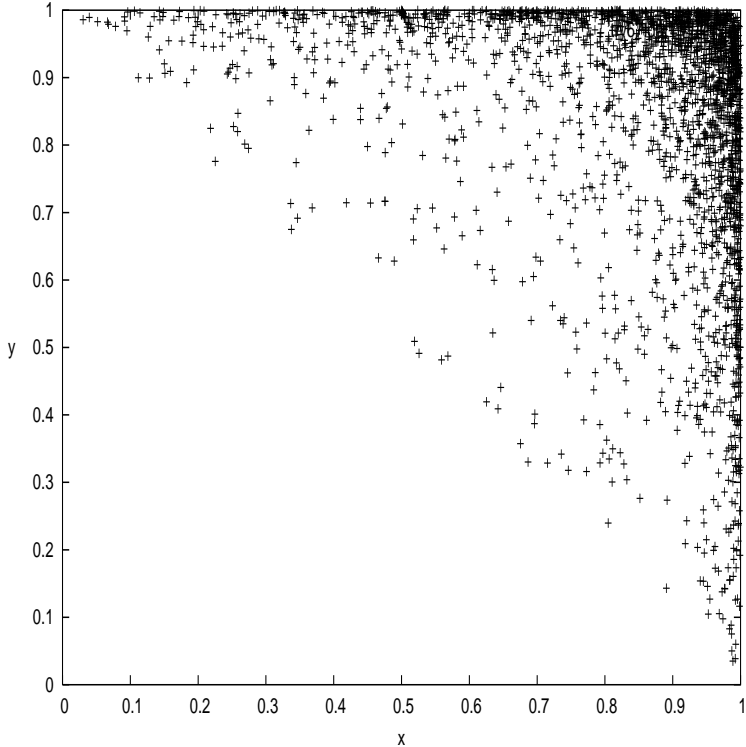,%
width=4in,height=4in,angle=270}
\end{center}
\caption{Phase space and distribution of hardest emissions.}
\label{fig:ps}
\end{figure}

To generate $x$ and $y$ values with a distribution proportional to
$U(x,y){\delta}(k_T(x,y)-p_T)$, we can use the $\delta$-function to eliminate the $y$
variable by computing
\begin{equation}
\centering
\label{eq:D}
D(x)=\int \delta(k_T-p_T)U(x,y)= \left.\frac{U(x,y)}{\frac{\partial k_T}{\partial y}}\right|_{y=\bar{y}}
\end{equation}
where $\bar{y}$ is such that $k_T(x,\bar{y})=p_T$. Note that $\frac{\partial k_T}{\partial
  y}$ is the same for both $y$ solutions. We then generate $x$ values with a
probability distribution proportional to $D$ with hit-and-miss techniques as described below. All events
generated have uniform weights. 
\begin{enumerate}[i)]
\item
Randomly sample $x$, $N_x$ times (we used $N_x=10^5$) in the range $[x_{\rm min}:x_{\rm max}]$ for the selected value of $\kappa$.\\
\item
For each value of $x$, evaluate $\bar{D}=D(x,y_1)+D(x,y_2)$ if there are two solutions for the selected $\kappa$ and
$\bar{D}=D(x,y_2)$ if there is only one solution. Also, if $\kappa<0.08333$ and $x<x^{u}$ (see
Figure \ref{fig:lt}), there is only one $y$ solution so evaluate $\bar{D}=D(x,y_2)$.\\
\item 
Find the maximum value $\bar{D}_{\rm max}$ of $\bar{D}$ for the selected value of $\kappa$ from the set of $N_x$ points that have been
sampled.\\
\item
Next, select a value for $x$ in the allowed range and evaluate $\bar{D}$.\\
\item
If  $\bar{D}> r\bar{D}_{\rm max}$ (where $r$ is a random number between $0$ and $1$),
accept the event, otherwise go to iv) and
generate a new value for $x$.\\
\item
If for the chosen value of $x$, there are two solutions for $y$, select a value for $y$ in
the ratio $D(x,y_1):D(x,y_2)$.
\item
Compare $U(x,y)$ with the true matrix element, $W(x,y)$. If the event fails this veto, set
$\kappa_{\rm max}=\kappa$ and
  regenerate a new $\kappa$ value as discussed in Section \ref{sec:xy}.\\
\end{enumerate}  
NB: For the region $y>x$, exchange $x$ and $y$ in the above discussion.
In this way, the smooth phase space distribution in Figure \ref{fig:ps} below was obtained
for the
hardest emission events. The plot show 2,500 of these events.
\section{Adding the truncated shower}
\label{sec:trunc}
As mentioned in Section \ref{sec:introduction}, a `truncated shower' would need to be added before the
hardest emission to simulate the soft radiation distribution. Due to angular ordering,
the `truncated' radiation  is emitted at a wider angle than the angle of the hardest emission as
well as a lower $p_T$. This means the `truncated' radiation does not appreciably degrade
the energy entering the hardest emission and justifies our decision to generate the
hardest emission first.

Below is an outline of how the `truncated shower' was generated. We will consider the case
in which at most one extra gluon is emitted by the quark or antiquark before the hardest
emission. The outline closely
follows the {\tt Herwig++} parton shower evolution method described in
\cite{Gieseke:2003hm, Gieseke:2003rz} where the
evolution variables $z$, the momentum fractions, and $\tilde{q}$, the evolution scale,
determine the kinematics of the shower. 
\begin{enumerate}[i)]
\item
Having generated the $p_T$ of the hardest emission and the energy fractions $x$ and $y$,
calculate the momentum fraction $z$ and $1-z$ of the partons involved in the hardest
emission. We will assume henceforth that
$x > y$ and that $y$ is the energy fraction of the quark, i.e. the quark is involved in the
hardest emission. Then
\begin{equation}
\centering
\label{eq:z}
z=\frac{y}{2-x}
\end{equation}
where as defined above $z$ and $1-z$ are the momentum fractions of the quark and gluon respectively. \\
\item
Next generate the momentum fraction $z_t$ of the `truncated' radiation according to the splitting function $P_{qq}=\frac{1+z^2}{1-z}$ within the range
\begin{equation}
\centering
\label{eq:mu}
\frac{\mu}{\tilde{q}_i}<z_t<1-\frac{Q_g}{\tilde{q}_i}
\end{equation}
where $\tilde{q}_i$ is the initial evolution scale, i.e. $\sqrt{s}=91.2$ GeV, $\mu$=max($m_a,Q_g$), $m_a$ is the mass of the quark and $Q_g$ is a cutoff
introduced to regularize soft gluon singularities in the splitting functions. In this
report, a $Q_g$ value of $0.75$ GeV (and hence $\mu=0.75$ GeV) was used. $z_t$ is the
momentum fraction of the quark after emitting the `truncated' gluon with momentum fraction $1-z_t$.\\
\item
Having generated a value for $z_t$, determine the scale $\tilde{q}_h$ of the hardest
emission from
\begin{equation}
\centering
\label{eq:q}
\tilde{q}_h=\sqrt{\frac{{p_T}^2}{z^2{(1-z)}^2}+\frac{\mu^2}{z^2}+\frac{{Q_g}^2}{z{(1-z)}^2}}
\end{equation}
where $z=z_t$.\\
\item
Starting from an initial scale $\tilde{q}_i$, the probability of there being an emission
next at the scale $\tilde{q}$ is given by
\begin{equation}
\centering
\label{eq:S}
S(\tilde{q}_i,\tilde{q})=\frac{\Delta(\tilde{q}_c,\tilde{q}_i)}{\Delta(\tilde{q}_c,\tilde{q})}
\end{equation}
where 
\begin{equation}
\centering
\label{eq:dt}
{\Delta(\tilde{q}_c,\tilde{q})}=exp\left[-\int_{\tilde{q}_c}^{\tilde{q}} 
  \frac{{d\tilde{q}}^{2}}{{\tilde{q}}^{2}}\int dz \frac{\alpha_s}{2\pi}P_{qq}\Theta(0< p_{T}^{t}< p_{T})\right].
\end{equation}
$\tilde{q}_c$ is the lower cutoff of the parton shower which was set to $0.4$ GeV in this
report, $\alpha_s$ is the running coupling constant evaluated at $z(1-z)\tilde{q}$,
$P_{qq}$ is the $q \rightarrow qg$ splitting function and
$k_T$ is the transverse momentum of the hardest emission. The Heaviside function ensures
that the transverse momentum, $ {p_{T}^{t}}$ of the truncated emission is real and
is less than $p_T$.
To evaluate the integral in (\ref{eq:dt}), we overestimate the integrands  and apply vetoes
with weights as described in
\cite{Gieseke:2003hm}. With $r$ a random number between $0$ and $1$, we then solve the equation 
\begin{equation}
\centering
\label{eq:r}
S(\tilde{q}_i,\tilde{q})=r
\end{equation}
for $\tilde{q}$. If $\tilde{q} > \tilde{q}_{h}$, the event has a `truncated'
emission. If $\tilde{q} < \tilde{q}_{h}$ , there is no `truncated' emission and the event is showered from the scale of the
hardest emission.\\
\item
If there is a `truncated' emission, the next step is to determine the transverse momentum
$p_{T}^{t}$ of the emission. This is given by \cite{Gieseke:2003hm}
\begin{equation}
\centering
\label{eq:pT}
p_{T}^{t}=\sqrt{(1-z_{t})^2({\tilde{q}}^{2}-\mu^2)-z_{t}{Q_{g}}^{2}}\;.
\end{equation}
If ${p_{T}^{t}}^2 < 0$ or  $p_{T}^{t} > p_{T}$ go to ii).\\
\item
We now have values for $z_t$, the momentum fraction of
the quark after the first emission, $p_{T}^{t}$, the transverse momentum of the
first emission, $z$, the momentum fraction of the hardest emission and $p_T$, the transverse momentum of the
hardest emission. This is illustrated in Figure \ref{fig:zpic}. We can then reconstruct the momenta of the partons as described in
\cite{Gieseke:2003hm}. The orientation of the quark, antiquark and hardest emission with
respect to the beam axis is determined as explained there for the hard matrix element correction.\\
\begin{figure}[!ht]
\begin{center}
\psfig{figure=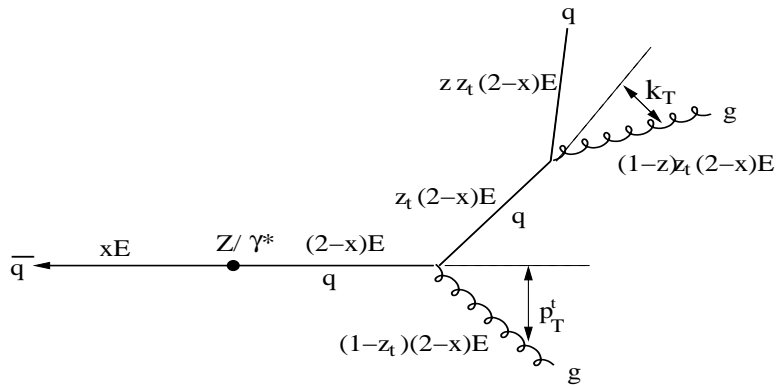,%
width=4in,height=2in,angle=0}
\end{center}
\caption{Adding the `truncated' emission. ($E=45.6$ GeV)}
\label{fig:zpic}
\end{figure}
\end{enumerate}

In principle this procedure could be iterated to generate a multi-gluon truncated
shower. However, for the present, we consider only the effect of at most one extra gluon
emission. As discussed in \cite{Nason:2004rx}, the initial showering scale of the hardest gluon is set
equal to the showering scale of the quark or antiquark closest in angle to it. This gives the right amount of soft radiation colour
connected to the gluon line.
\section{Results and data comparisons}
\label{sec:results}

\begin{figure}[!ht]
\begin{center}
\psfig{figure=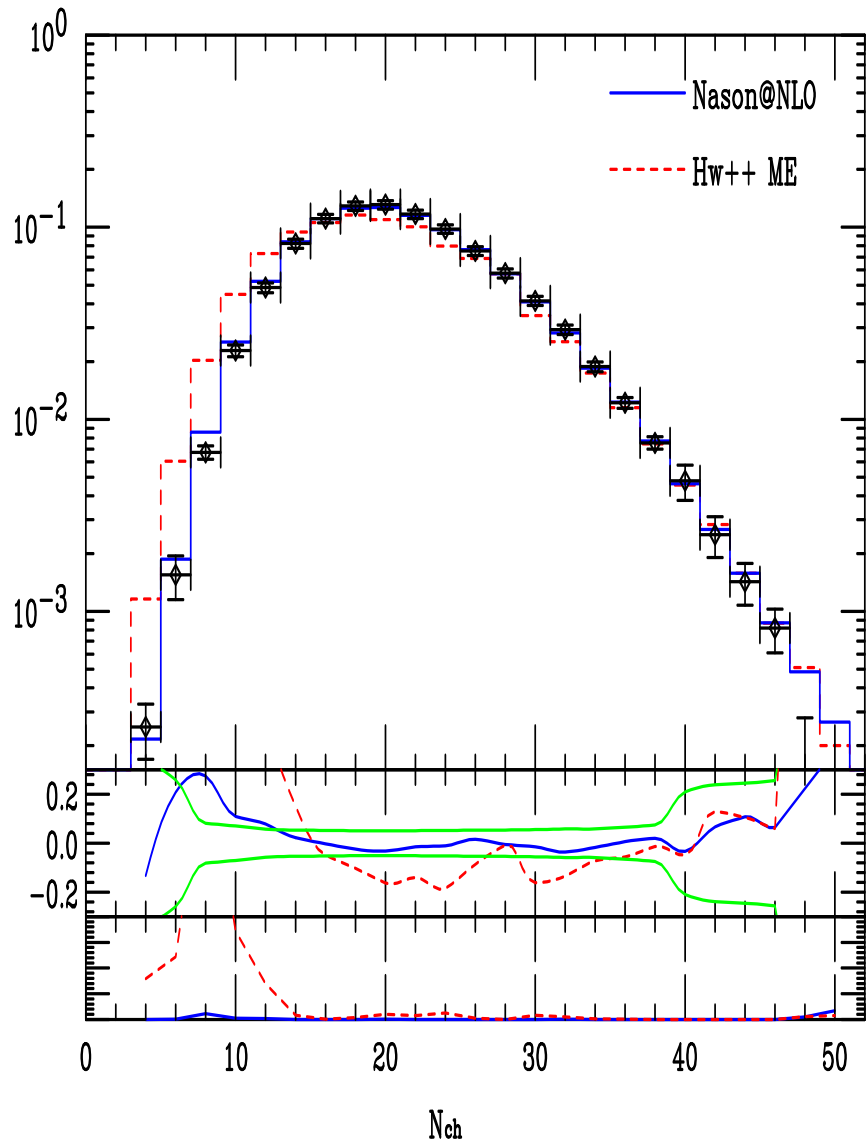,%
width=6in,height=4.5in,angle=90}
\end{center}
\caption{The distribution of the charged particle multiplicity. Data from the OPAL experiment at LEP \cite{Acton:1991aa}.}
\label{fig:mt}    
\end{figure}

One million events were generated as described above and then interfaced with the SMC,
{\tt Herwig++}. $13\%$ of the events acquired an extra `truncated' gluon. A $p_T$ veto
was imposed on the subsequent shower starting from the hardest emission to the
hadronization scale which was tuned to $0.4$ GeV.  Table \ref{tab:mult} and Figure \ref{fig:mt} show respectively the average multiplicities of a wide range of hadron
species and the charged particle multiplicity distribution. The subsequent figures are
plots of comparisons with event shape distributions from the DELPHI experiment at
LEP \cite{Abreu:1996na}. 

The upper panel below the main histograms shows the ratio
$\frac{M_{i}-D_{i}}{D_{i}}$ (where $M_i$ and $D_i$ stand for Monte Carlo result and data
value respectively) compared with the relative experimental error (green). The lower panel
shows the relative contribution to the $\chi^2$ of each observable. As in \cite{Gieseke:2003hm}, the
$\chi^2$ contributions allow for a $5\%$ uncertainty in the predictions if the data are
more accurate than this.
Finally, in Table 2 we show a list of $\chi^2$ values for all observables that were
studied during the analysis. The results were generated using
{\tt Herwig++} $2.0$ \cite{Gieseke:2006ga} which includes some improvements in the simulation of the
shower and hadronization as described in \cite{Gieseke:2006ga} leading to some changes in the  $\chi^2$
for specific observables relative to  \cite{Gieseke:2003hm}, although in most cases the changes are small.

\section{Conclusions}
\label{sec:conclusions}
We have successfully implemented the Nason method of generating the hardest emission first
and subsequently adding a truncated shower for $e^+e^-$ annihilation into hadrons. 

We have tested the method against data from $e^+e^-$ colliders and for almost all examined
observables, the simulation of the data
is improved with respect to {\tt Herwig++}. In particular the Nason method seems to fit
the data better in the soft regions of phase space. The poorer fits obtained for variables such as the thrust minor which
vanish in the three-jet limit (and in general for planar events) may be attributable to the lack of multiple emission in the truncated shower.  

The fits are summarized in the $\chi^2$ values shown in Table \ref{tab:chi2}. In Table
\ref{tab:chi2} we also present the $\chi^2$ values for the
observables obtained without the implementation of the `truncated' shower. We see there is
a general improvement in the fits associated with the addition of the `truncated' emissions. 

Future work in this area will extend this method to processes with initial state radiation in
hadron-hadron collisions aiming at simulating Tevatron and LHC events.

\section*{Acknowledgements}
We are most grateful to all the other members of the {\tt Herwig++} collaboration for
developing the program that underlies the present work. This work was supported in part by
the UK Particle Physics and Astronomy Research Council.
 
\clearpage
\begin{table}
\small
\begin{center}
\begin{tabular}{llllll} \hline
Particle & Experiment & Measured & {\tt Herwig++} ME & Nason@NLO
\\
\hline
\hline
All Charged & M,A,D,L,O & 20.924 $\pm$ 0.117  & $20.9157$   & $21.117$     \\
\hline                                                                    
$\gamma$  & A,O & 21.27 $ \pm$ 0.6 & $23.02$                & $22.80$    \\ 
$\pi^0$   & A,D,L,O & 9.59 $ \pm$ 0.33 & $10.44$            & $10.45$    \\ 
$\rho(770)^0$ & A,D & 1.295 $ \pm$ 0.125 & $1.201$          & $1.19$     \\ 
$\pi^\pm$ & A,O & 17.04 $ \pm$ 0.25 & $17.28$               & $17.36$     \\ 
$\rho(770)^\pm$ & O & 2.4 $ \pm$ 0.43 & $1.94$              & $1.85$      \\ 
$\eta$ & A,L,O & 0.956 $ \pm$ 0.049 & $0.976$               & $1.02$  \\ 
$\omega(782)$ & A,L,O & 1.083 $ \pm$ 0.088& $0.851$         & $0.745$    \\ 
$\eta'(958)$ & A,L,O & 0.152 $ \pm$ 0.03 & $0.138$           & $0.084$   \\ 
\hline                                                                     
$K^0$ & S,A,D,L,O & 2.027 $ \pm$ 0.025 & $1.949^*$          & $2.742^*$    \\ 
$K^*(892)^0$ & A,D,O & 0.761 $ \pm$ 0.032 & $0.627^*$         & $0.713$  \\ 
$K^*(1430)^0$ & D,O & 0.106 $ \pm$ 0.06 & $0.084$           & $0.033$   \\ 
$K^\pm$ & A,D,O & 2.319 $ \pm$ 0.079 & $2.155$              & $2.259$    \\ 
$K^*(892)^\pm$ & A,D,O & 0.731 $ \pm$ 0.058 & $0.614$       & $0.699$   \\ 
$\phi(1020)$ & A,D,O & 0.097 $ \pm$ 0.007 & $0.116$         & $0.101$   \\ 
\hline                                                                     
$p$ & A,D,O & 0.991 $ \pm$ 0.054 & $0.871$                  & $0.958$    \\ 
$\Delta^{++}$ & D,O & 0.088 $ \pm$ 0.034 & $0.085$          & $0.086$  \\ 
$\Sigma^-$ & O & 0.083 $ \pm$ 0.011 & $0.0711$               & $0.073$   \\ 
$\Lambda$ & A,D,L,O & 0.373 $ \pm$ 0.008 & $0.391$        & $0.030$  \\ 
$\Sigma^0$ & A,D,O & 0.074 $ \pm$ 0.009 & $0.091$           & $0.097$   \\ 
$\Sigma^+$ & O & 0.099 $ \pm$ 0.015 & $0.077$               & $0.089$   \\ 
$\Sigma(1385)^\pm$ & A,D,O & 0.0471 $ \pm$ 0.0046 & $0.0307^*$ & $0.031$ \\
$\Xi^-$ & A,D,O & 0.0262 $ \pm$ 0.001 & $0.032^*$             & $0.034^*$   \\ 
$\Xi(1530)^0$ & A,D,O & 0.0058 $ \pm$ 0.001 & $0.0092^*$     & $0.0096^*$ \\ 
$\Omega^-$ & A,D,O & 0.00125 $ \pm$ 0.00024 & $0.00197$       & $0.00194$  \\ 
\hline                                                                     
$f_2(1270)$ & D,L,O & 0.168 $ \pm$ 0.021 & $0.164$          & $0.173$    \\ 
$f_2'(1525)$ & D & 0.02 $ \pm$ 0.008 & $0.0146$              & $0.0184$   \\ 
$D^\pm$ & A,D,O & 0.184 $ \pm$ 0.018 & $0.252^*$            & $0.282^*$ \\ 
$D^*(2010)^\pm$ & A,D,O & 0.182 $ \pm$ 0.009 & $0.173$      & $0.185$   \\ 
$D^0$ & A,D,O & 0.473 $ \pm$ 0.026 & $0.497$              & $0.497$ \\ 
$D_s^\pm$ & A,O & 0.129 $ \pm$ 0.013 & $0.130$            & $0.150$ \\ 
$D_s^{*\pm}$ & O & 0.096 $ \pm$ 0.046 & $0.06$             & $0.044$   \\ 
$J/\Psi$ & A,D,L,O & 0.00544 $ \pm$ 0.00029 & $0.00349^*$       & $0.0033^*$\\ 
$\Lambda_c^+$ & D,O & 0.077 $ \pm$ 0.016 & $0.0215^*$        & $0.0303^*$ \\ 
$\Psi'(3685)$ & D,L,O & 0.00229 $ \pm$ 0.00041 & $0.00165$  & $0.00154$  \\ 
\hline
\hline
\label{tab:mult}
\end{tabular}
\end{center}
\caption{Multiplicities per event at $91.2$ GeV. We show results 
  from {\tt Herwig++} with a matrix element correction (Herwig++ ME) and the
  new method, Nason@NLO . Experiments
  are Aleph(A), Delphi(D), L3(L), Opal(O), Mk2(M) and SLD(S). The $*$
  indicates a prediction that differs from the measured value by
  more than three standard deviations. Data are combined and updated from a variety of
  sources, see ref \cite{Webber:1999ui}.
 } 
\end{table}
\pagebreak
\vspace{4cm}

\pagebreak
\begin{figure}[!ht]
\vspace{2cm}
\hspace{0.5cm}
\psfig{figure=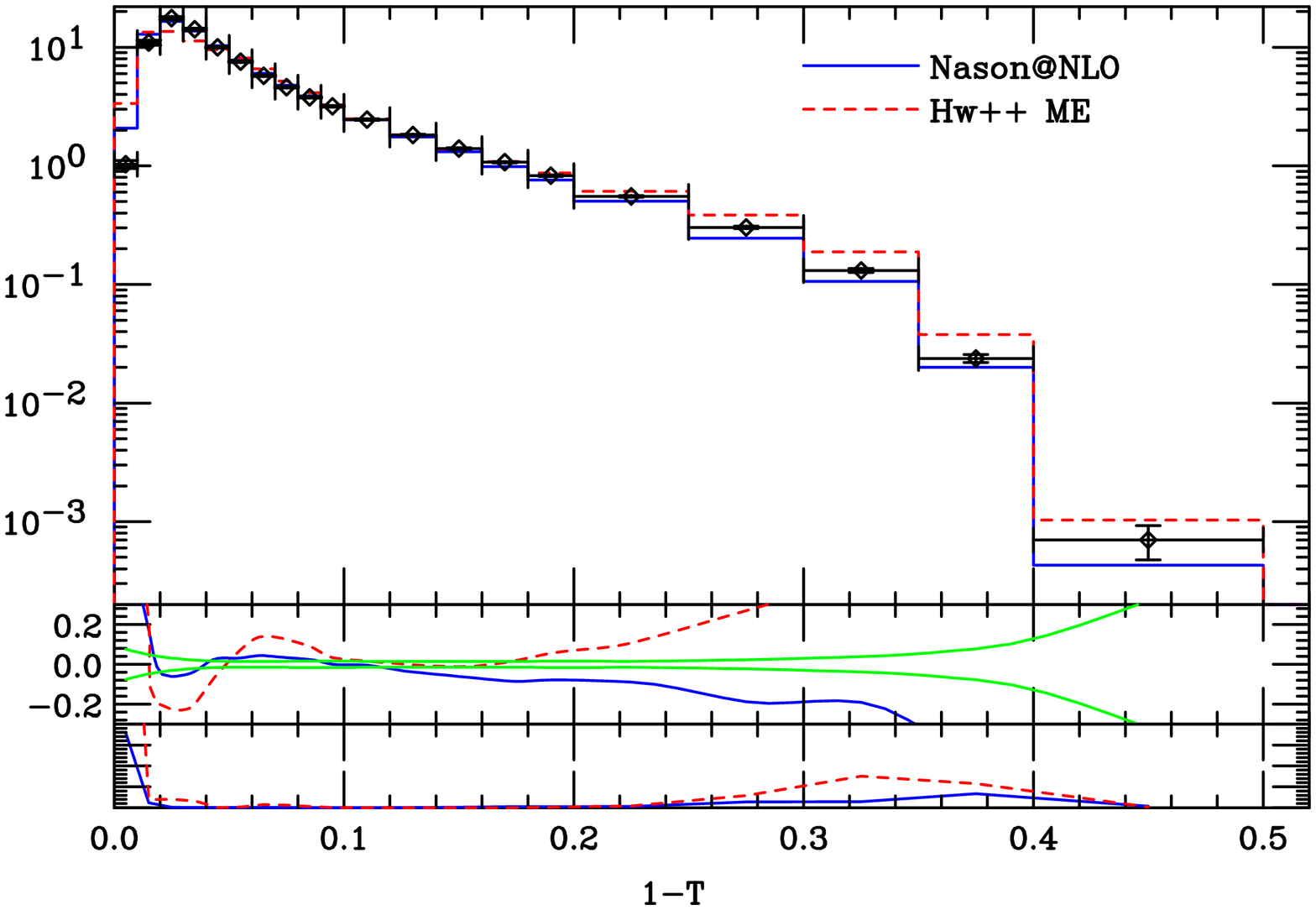,%
width=3.63in,height=2.63in,angle=90}
\psfig{figure=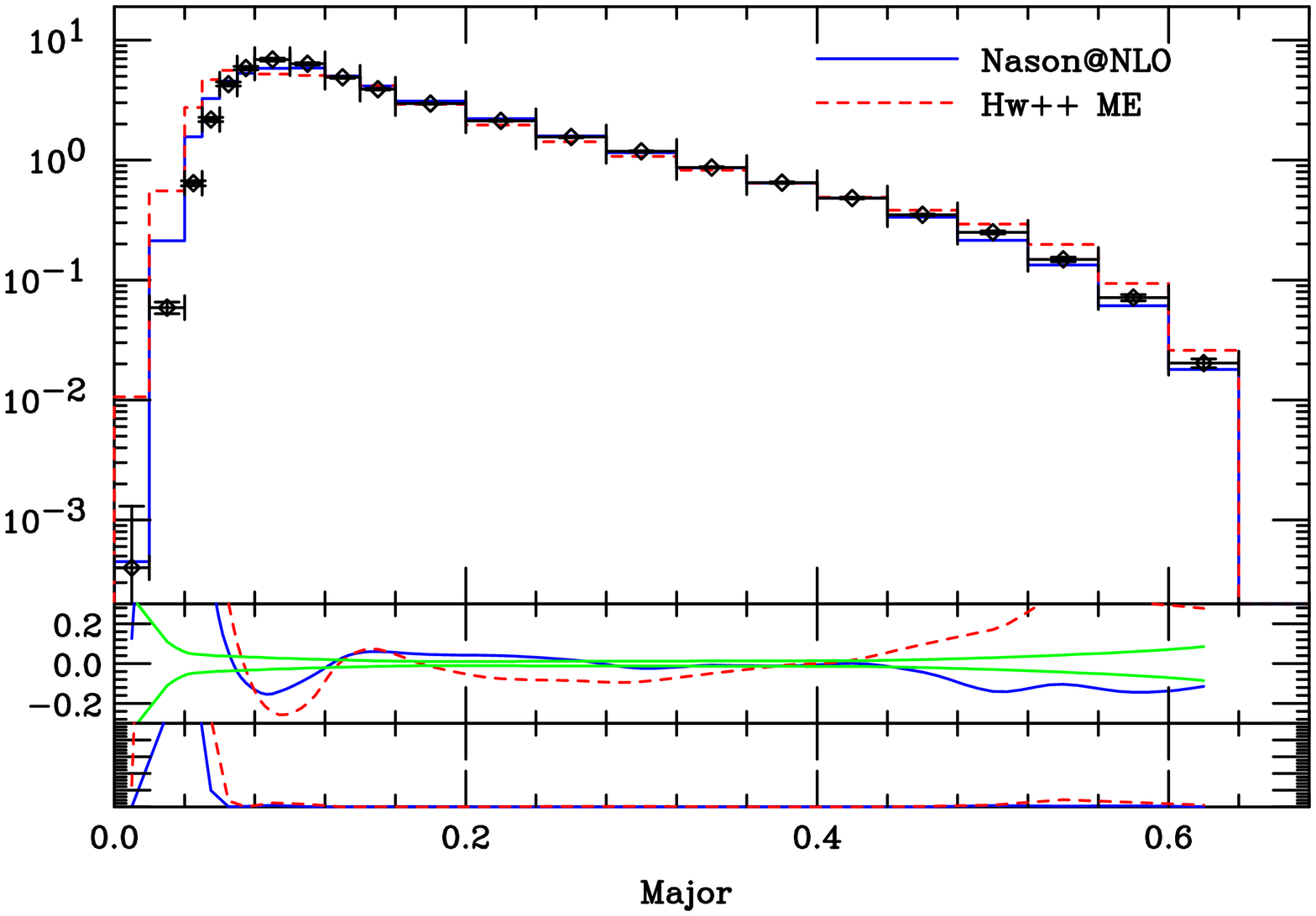,%
width=3.63in,height=2.63in,angle=90} 

\hspace{4cm}
\psfig{figure=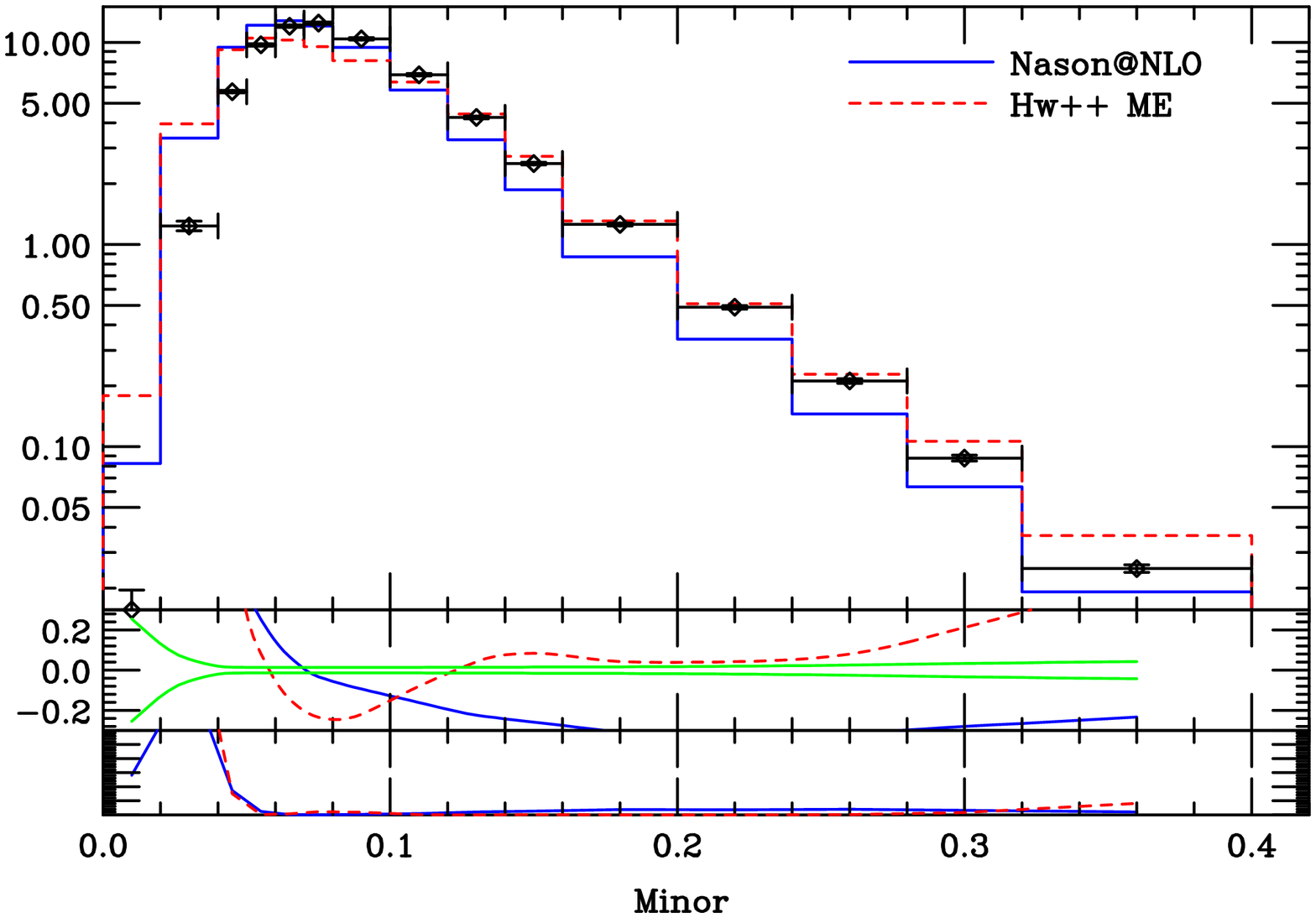,%
width=3.63in,height=2.63in,angle=90}
\caption{Thrust, Thrust major and Thrust minor. Data from the DELPHI experiment at LEP \cite{Abreu:1996na}.}
\label{fig:tt}    
\end{figure}
\pagebreak
\begin{figure}[!ht]
\vspace{2cm}
\hspace{0.5cm}
\psfig{figure=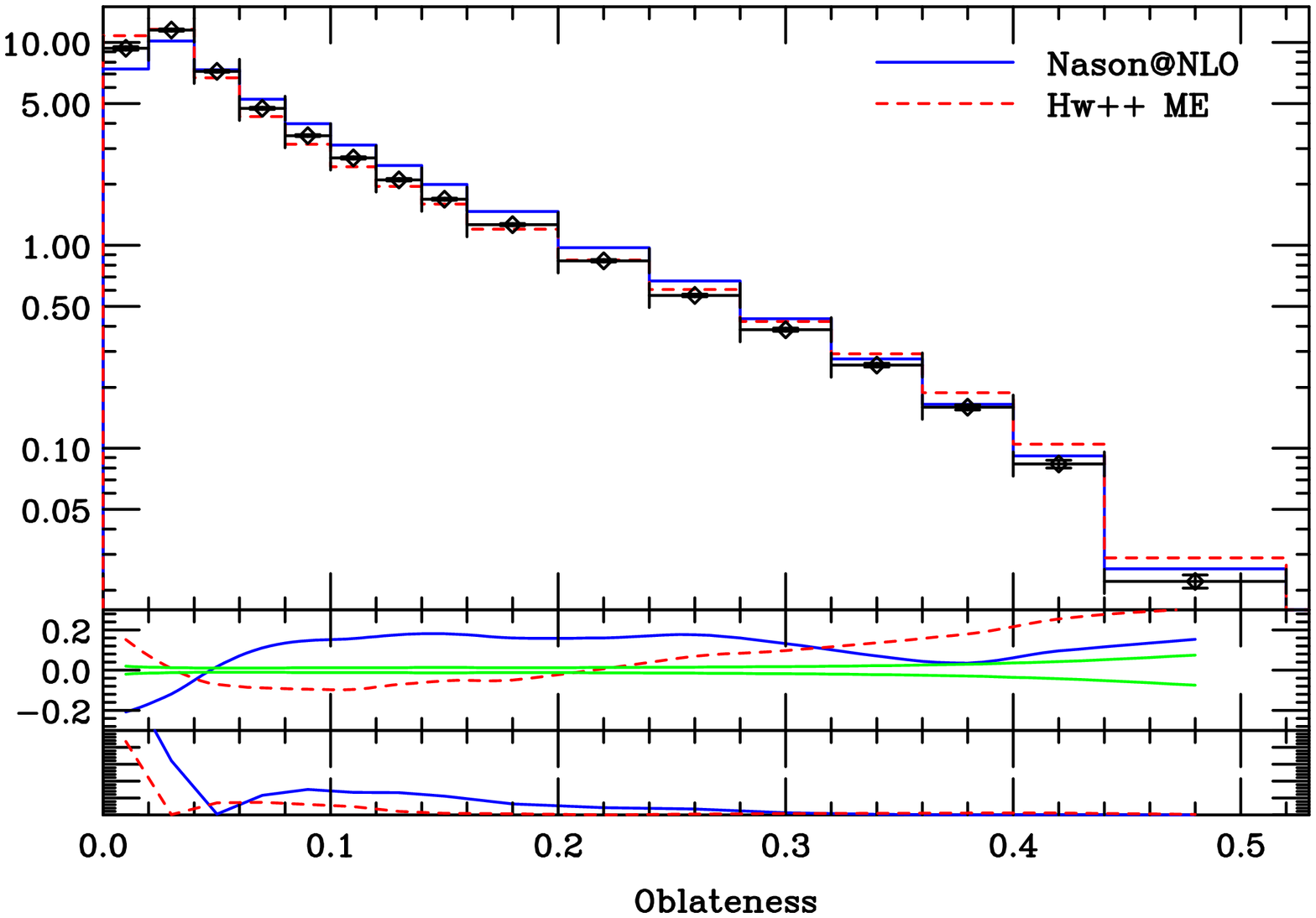,%
width=3.63in,height=2.63in,angle=90}
\psfig{figure=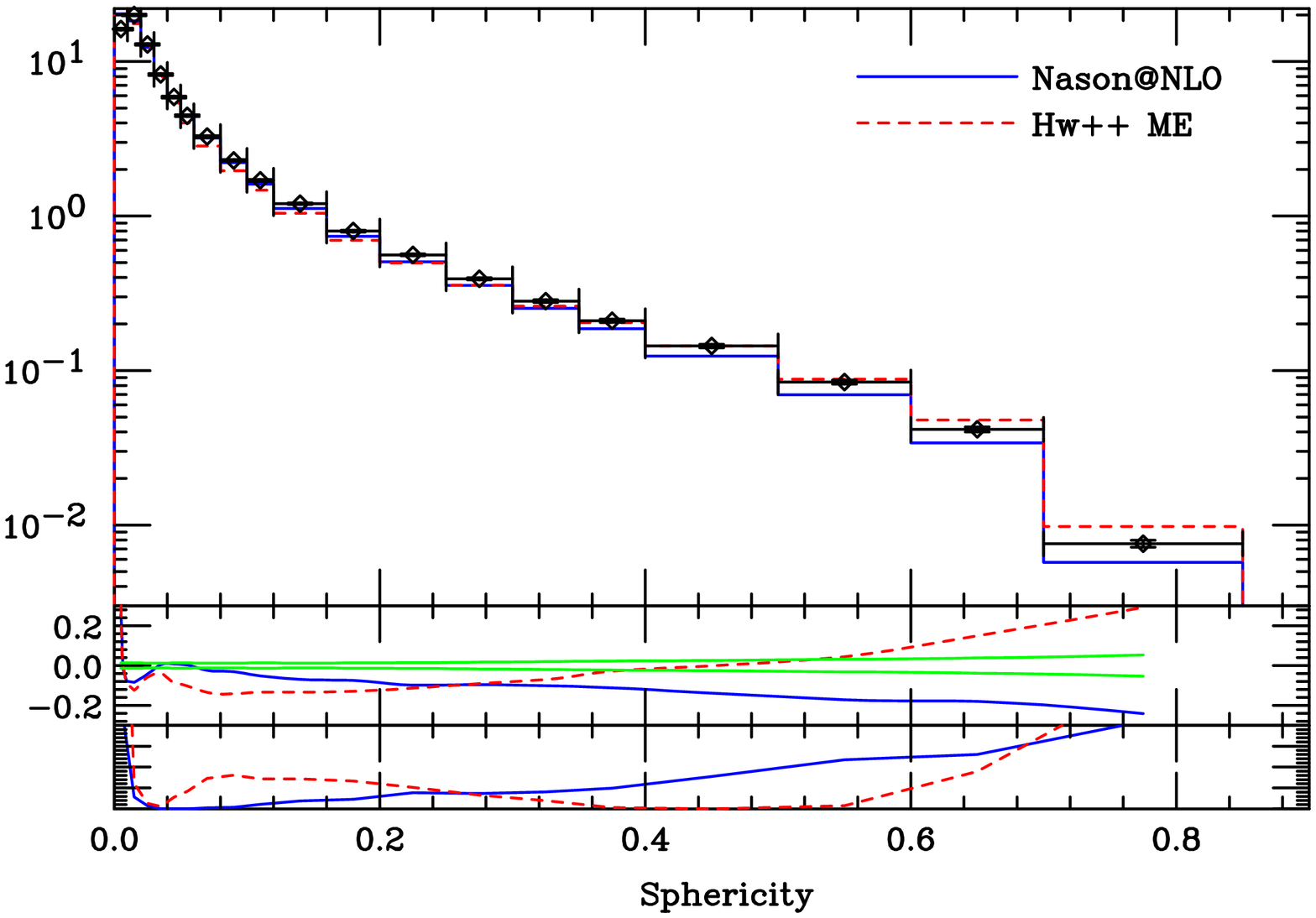,%
width=3.63in,height=2.63in,angle=90} 

\hspace{0.5cm}
\psfig{figure=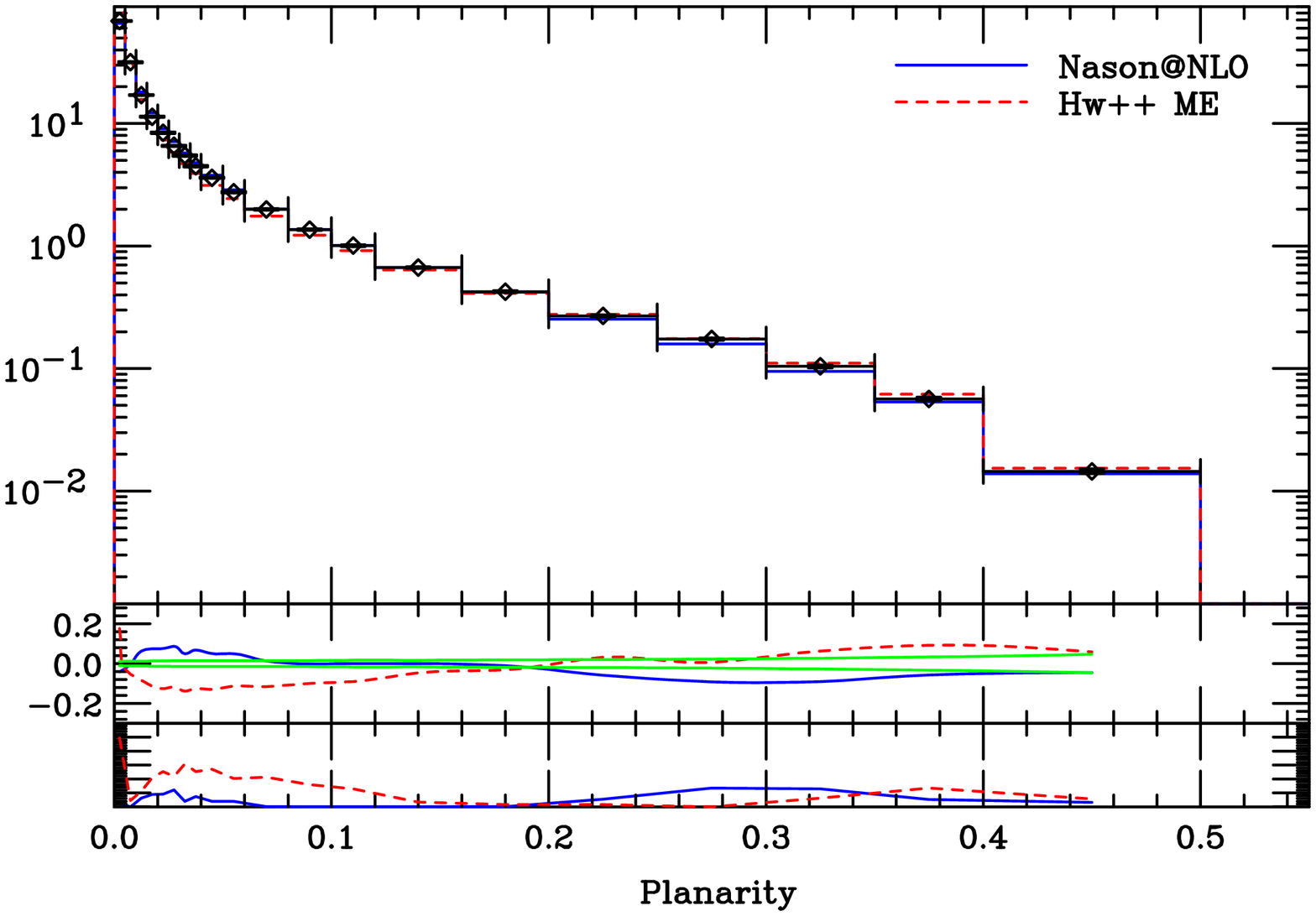,%
width=3.63in,height=2.63in,angle=90}
\psfig{figure=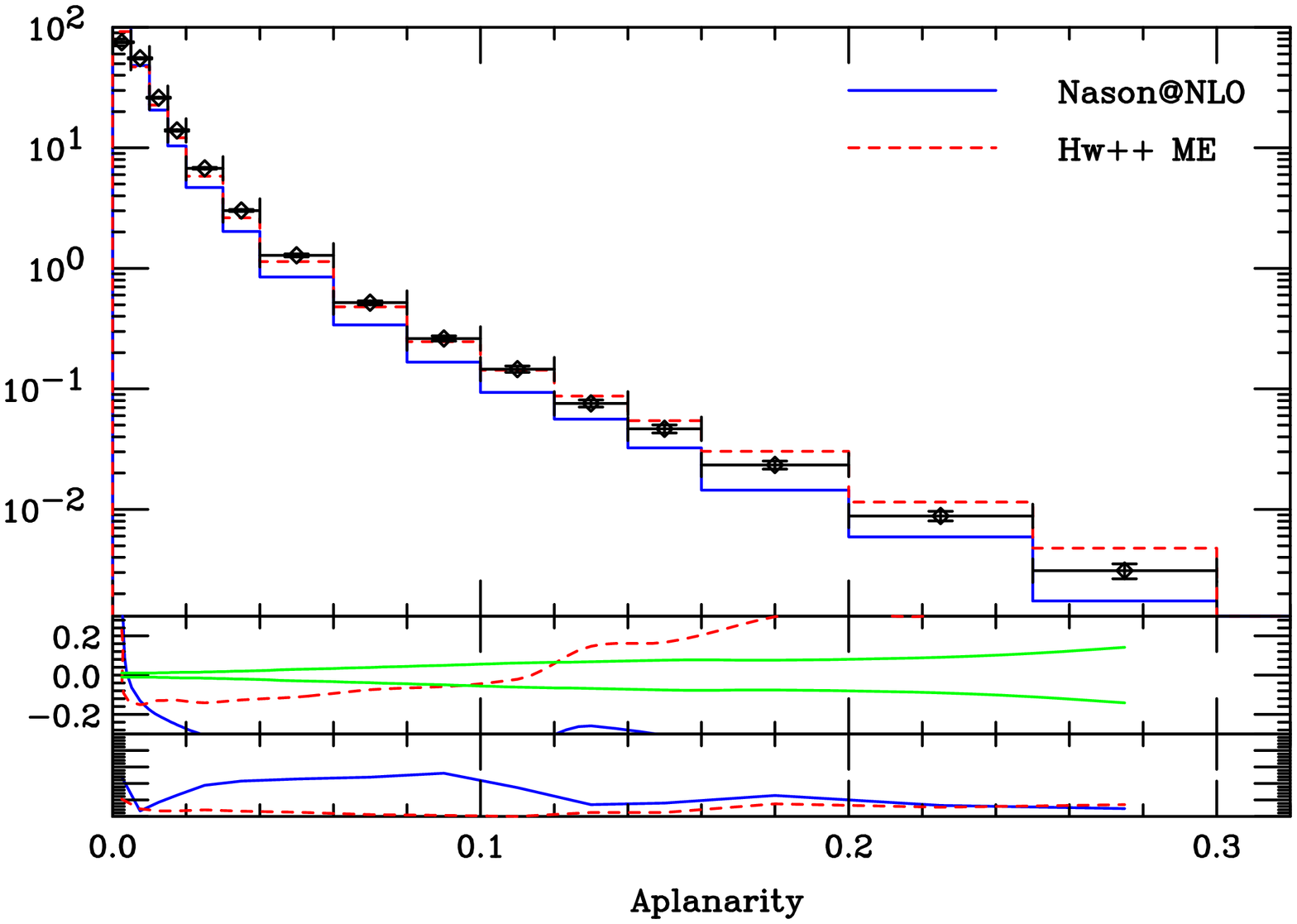,%
width=3.63in,height=2.63in,angle=90}
\caption{Oblateness, Sphericity, Aplanarity and Planarity distributions. Data from the DELPHI experiment at LEP \cite{Abreu:1996na}.}
\label{fig:osap}    
\end{figure}
\pagebreak
\begin{figure}[!ht]
\vspace{2cm}
\hspace{0.5cm}
\psfig{figure=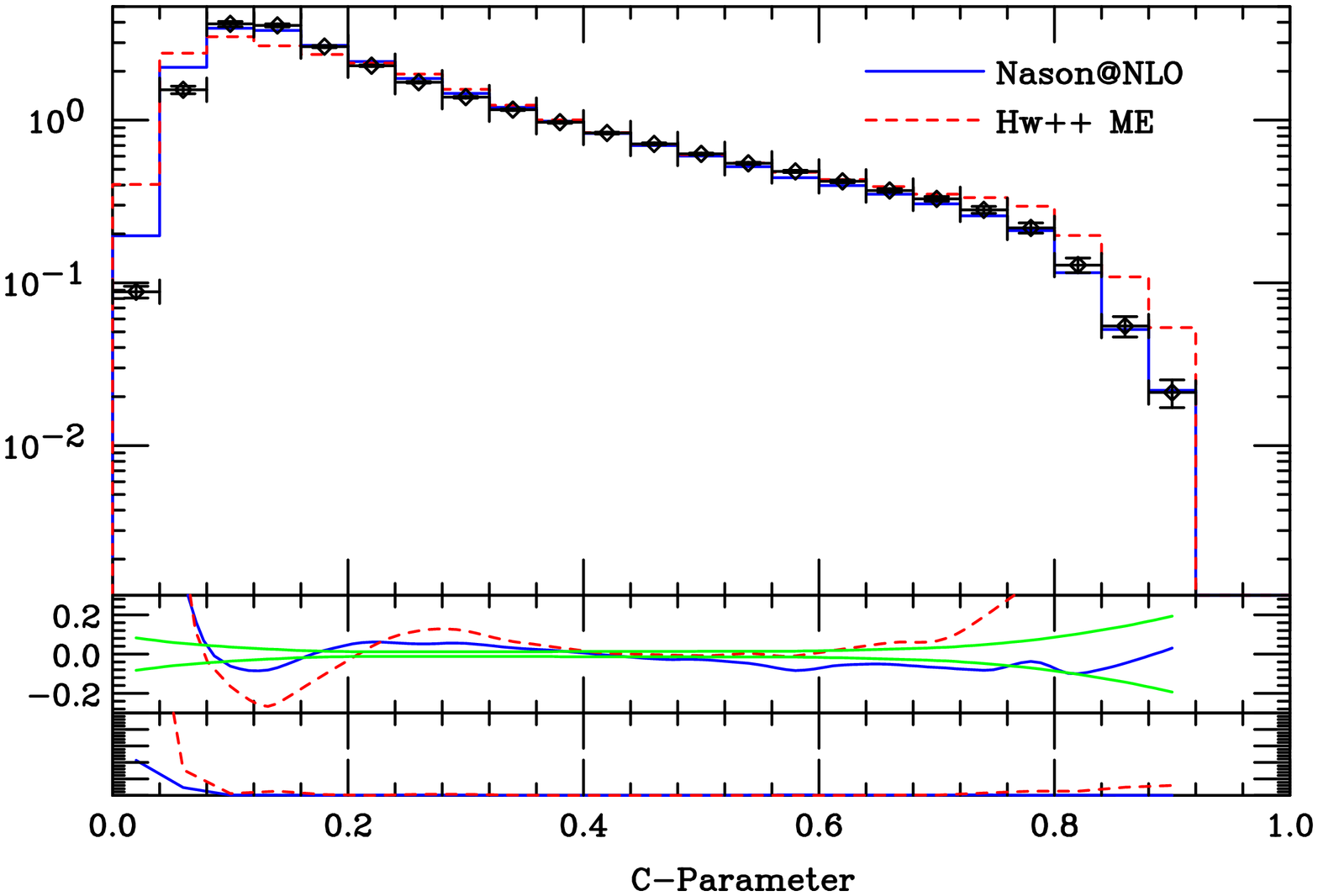,%
width=3.63in,height=2.63in,angle=90}
\psfig{figure=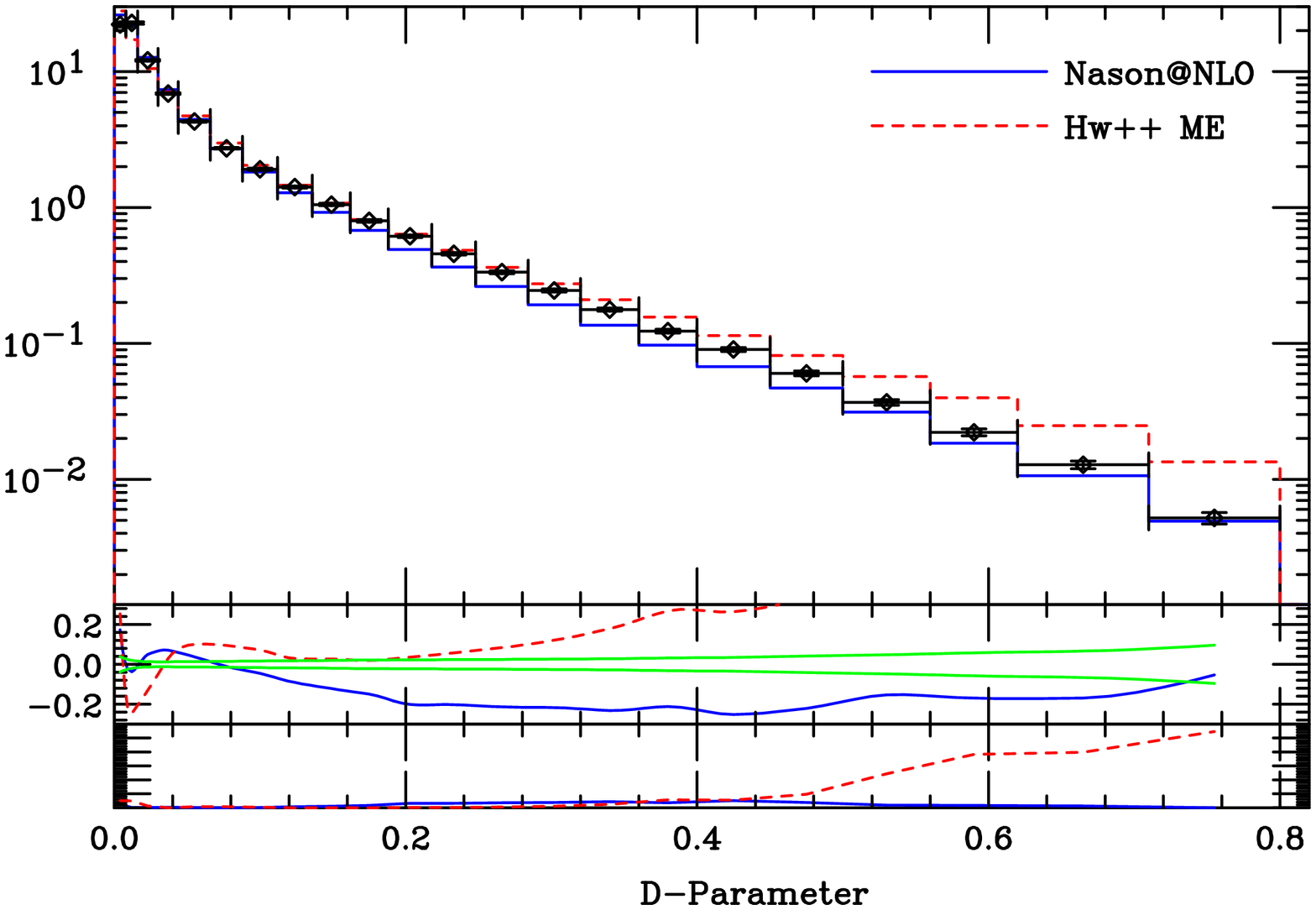,%
width=3.63in,height=2.63in,angle=90} 

\hspace{0.5cm}
\psfig{figure=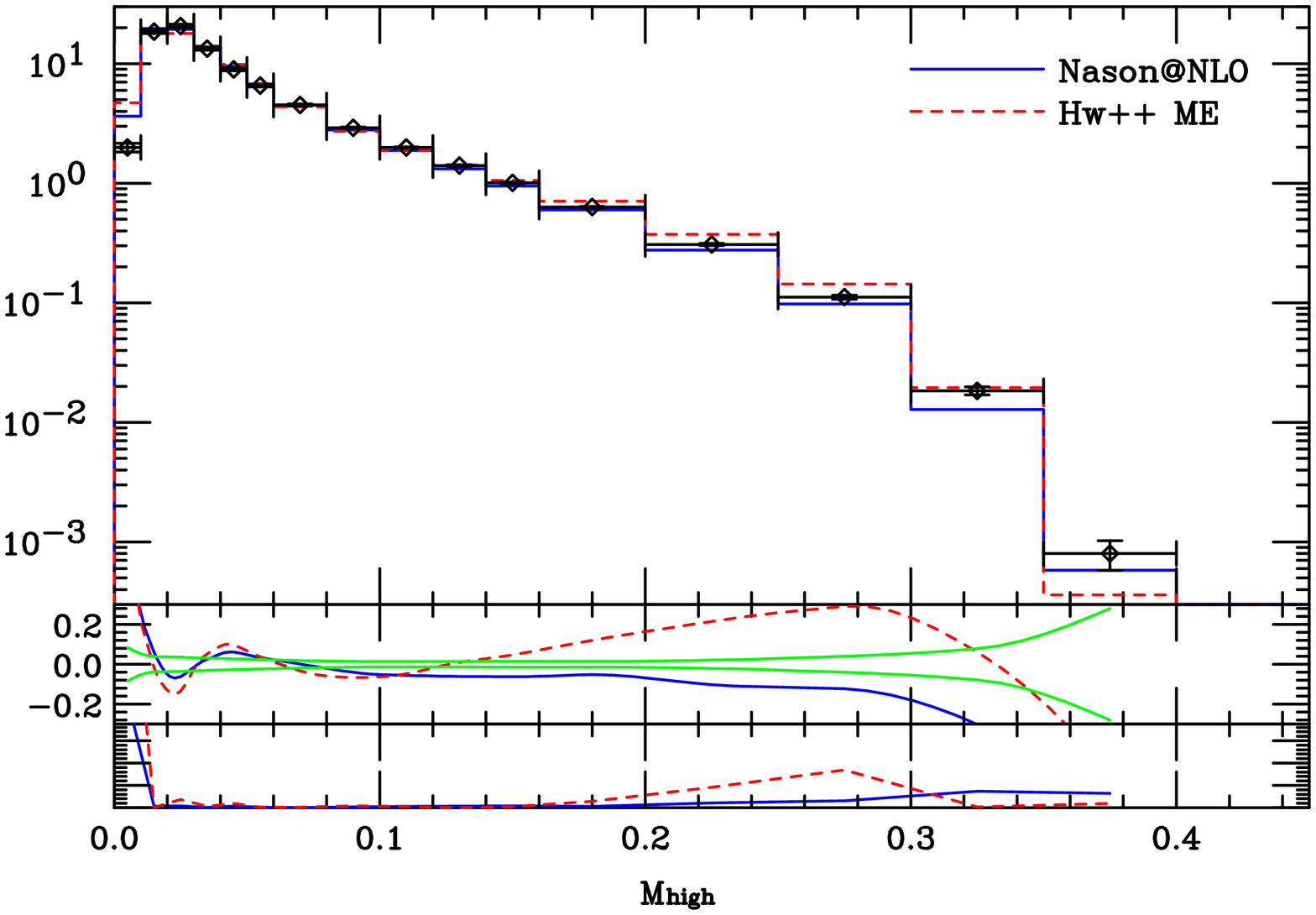,%
width=3.63in,height=2.63in,angle=90}
\psfig{figure=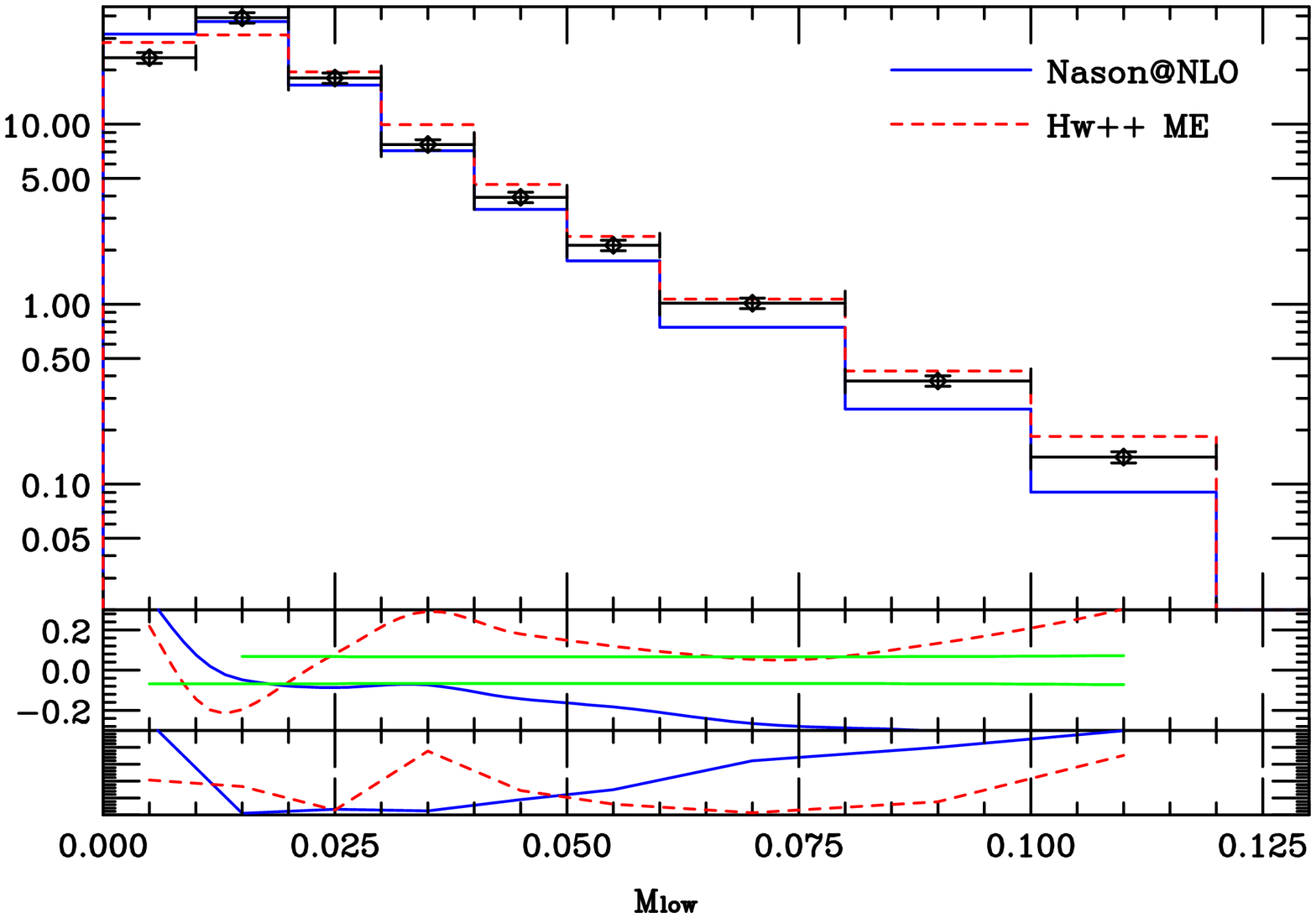,%
width=3.63in,height=2.63in,angle=90}
\caption{$C$ Parameter and $D$ Parameter distributions and the high, $M_{\rm high}$ and
  low, $M_{\rm low}$  hemisphere masses. Data from the DELPHI experiment at LEP \cite{Abreu:1996na}.}
\label{fig:cdlh}    
\end{figure}
\pagebreak
\begin{figure}[!ht]
\vspace{2cm}
\hspace{0.5cm}
\psfig{figure=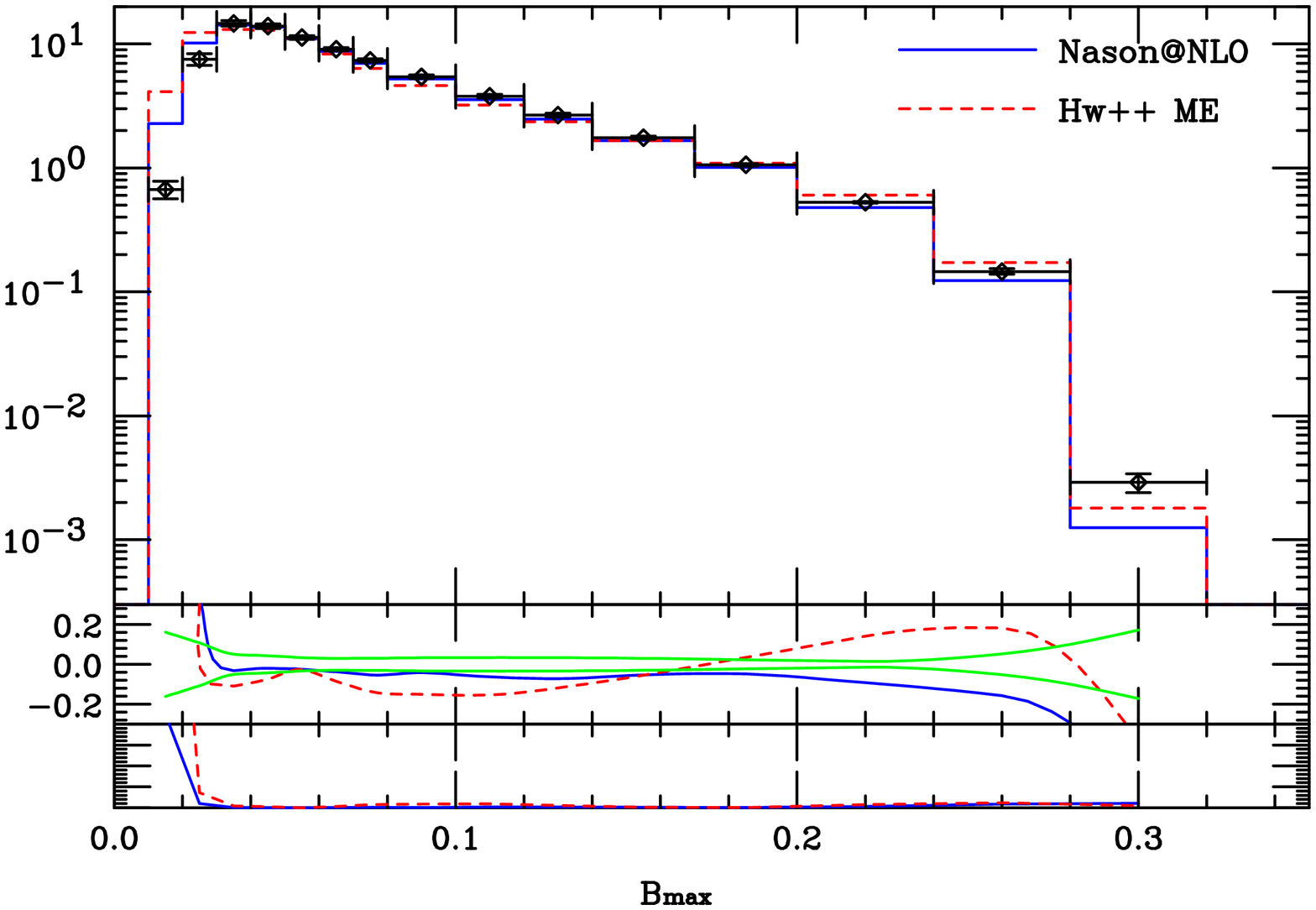,%
width=3.63in,height=2.63in,angle=90}
\psfig{figure=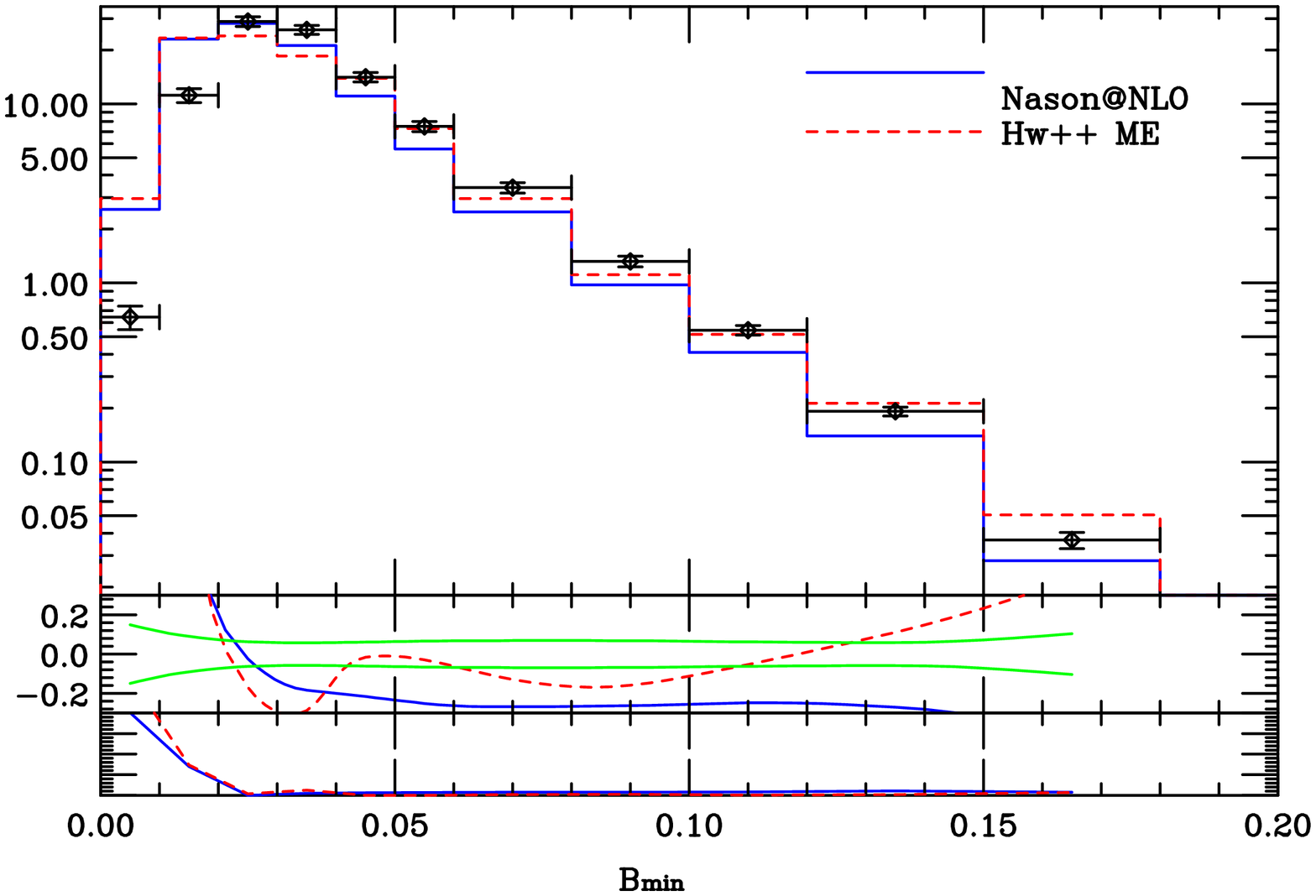,%
width=3.63in,height=2.63in,angle=90} 

\hspace{4cm}
\psfig{figure=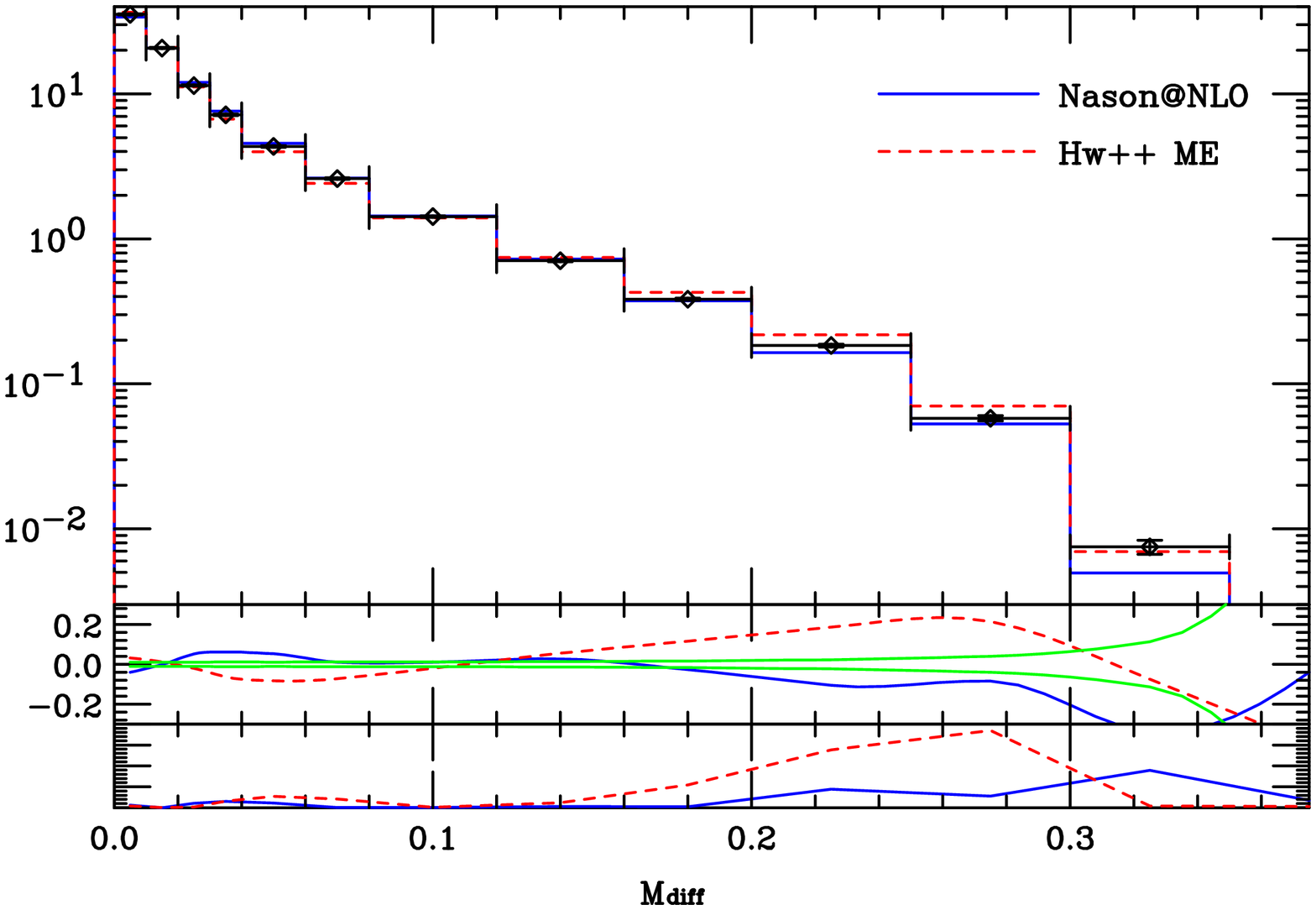,%
width=3.63in,height=2.63in,angle=90}
\caption{The wide and narrow jet broadening measures $B_{\rm max}$ and $B_{\rm min}$ and
  the difference in hemisphere masses, $M_{\rm diff}$. Data from the DELPHI experiment at
  LEP \cite{Abreu:1996na}.}
\label{fig:xnmd} 
\pagebreak   
\end{figure}
\begin{figure}[!ht]
\hspace{0.5cm}
\psfig{figure=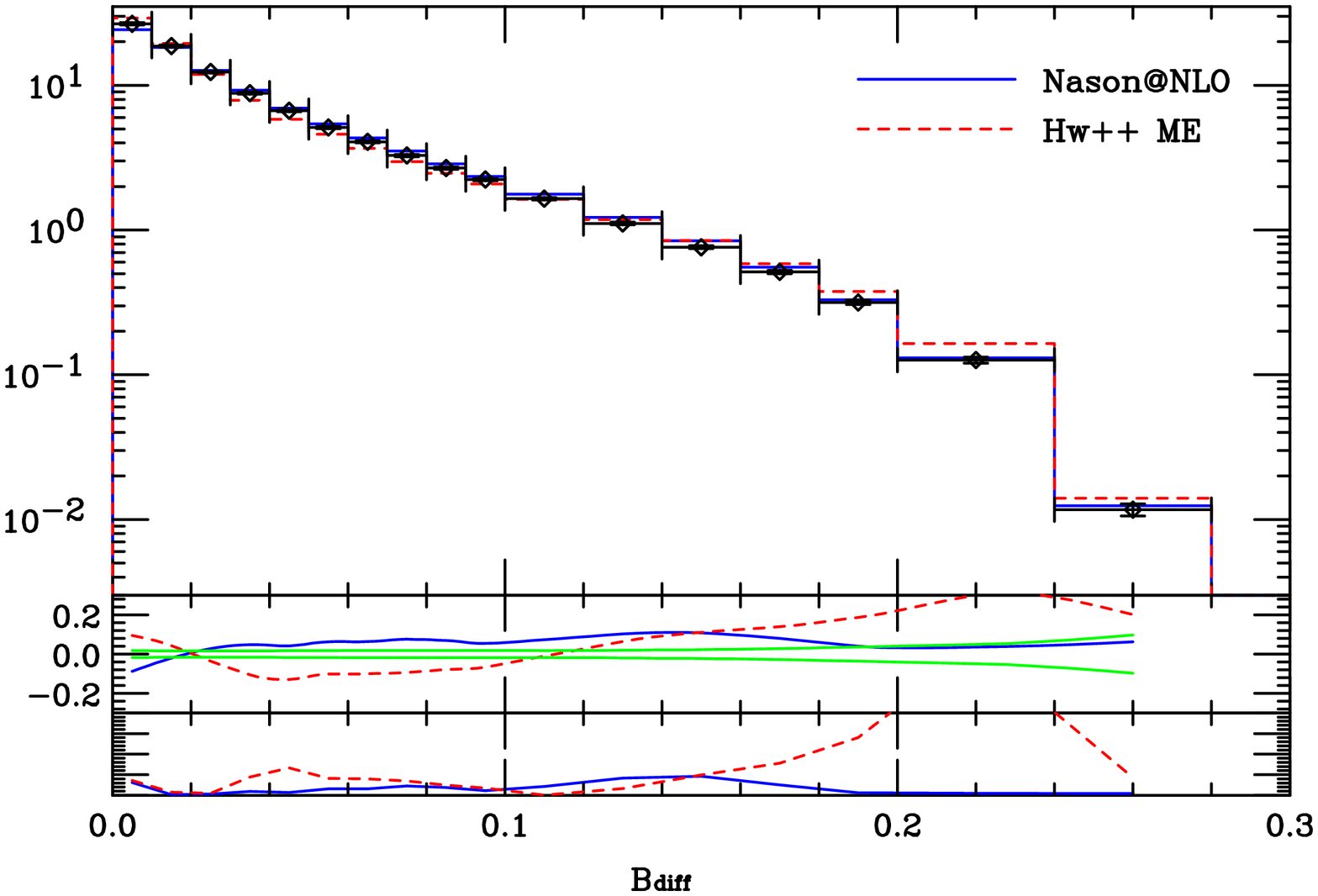,%
width=3.63in,height=2.63in,angle=90}
\psfig{figure=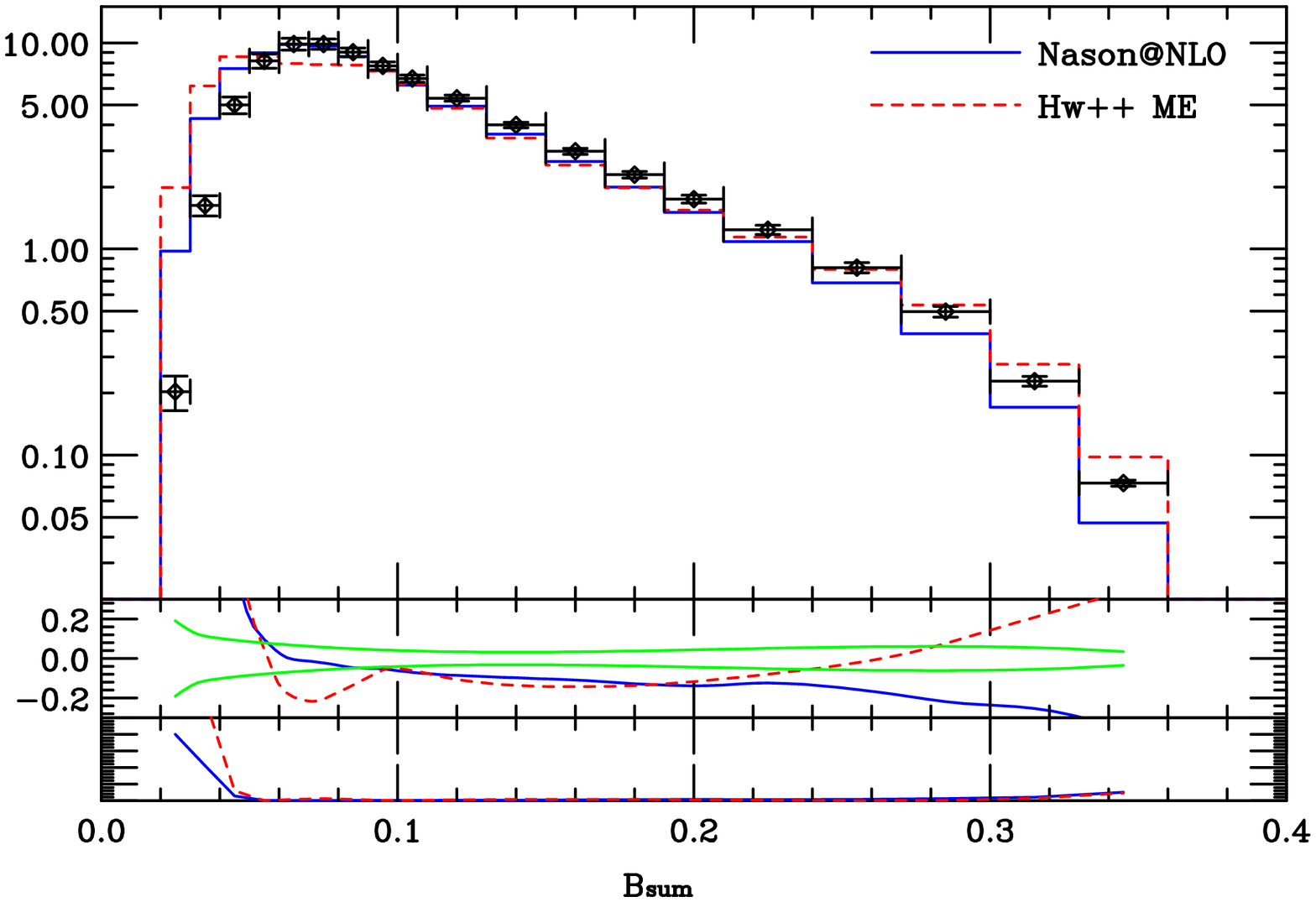,%
width=3.63in,height=2.63in,angle=90} 
\caption{The difference and sum of jet broadenings $B_{\rm diff}$ and $B_{\rm sum}$. Data
  from the DELPHI experiment at LEP \cite{Abreu:1996na}.}
\label{fig:bsbd}    
\end{figure}

\clearpage
\vspace{4cm}
\begin{table}
\centering
\begin{tabular}{cccc}
\hline
\hline
Observable &{\tt Herwig++} ME&Nason@NLO&Nason@NLO\\
           &                 & with truncated shower& w/o truncated shower \\
\hline
\hline
$1-T$                         &36.52& 9.03&9.81 \\
Thrust Major                &267.22&36.44&37.65 \\
Thrust Minor                &190.25&86.30&90.59 \\
Oblateness                  &  7.58&  6.86&6.28 \\
Sphericity                  &  9.61&  7.55&9.01 \\
Aplanarity                   &  8.70&  22.96&25.33 \\
Planarity                  & 2.14& 1.19&1.45 \\
$C$ Parameter                 & 96.69& 10.50&11.14 \\
$D$ Parameter                 & 84.86& 8.89&10.88 \\
$M_{\rm high}$                 & 14.70& 5.31&6.61 \\
$M_{\rm low}$                  & 7.82&  12.90&13.44 \\
$M_{\rm diff}$                 &  5.11&  1.89&2.09 \\
$B_{\rm max}$                  & 39.50& 11.42&12.17 \\
$B_{\rm min}$                  & 45.96& 35.2&36.16 \\
$B_{\rm sum}$                  &91.03& 28.83&30.58\\
$B_{\rm diff}$                 &  8.94&  1.40&1.14 \\
$N_{ch}$                      &43.33&1.58&10.08\\
\hline
\hline
$\langle\chi^2\rangle/{\rm bin}$& 56.47& 16.96&18.49\\
\hline
\hline
\end{tabular}
\vspace{0.5cm}
\caption{$\chi^2/$bin for all observables we studied.}
\label{tab:chi2}
\end{table}

\clearpage
\bibliography{ZNason}

\providecommand{\href}[2]{#2}\begingroup\raggedright\begin{thebibliography}{10}

\bibitem{Frixione:2002ik}
S.~Frixione and B.~R. Webber, ``Matching {NLO QCD} computations and parton
  shower simulations,'' {\em JHEP} {\bf 06} (2002) 029,
\href{http://www.arXiv.org/abs/hep-ph/0204244}{{\tt hep-ph/0204244}}.

\bibitem{Frixione:2003ei}
S.~Frixione, P.~Nason, and B.~R. Webber, ``Matching {NLO QCD} and parton
  showers in heavy flavour production,'' {\em JHEP} {\bf 08} (2003) 007,
\href{http://www.arXiv.org/abs/hep-ph/0305252}{{\tt hep-ph/0305252}}.

\bibitem{Frixione:2006gn}
S.~Frixione and B.~R. Webber, ``The {MC@NLO} 3.3 event generator,''
\href{http://www.arXiv.org/abs/hep-ph/0612272}{{\tt hep-ph/0612272}}.

\bibitem{Nason:2004rx}
P.~Nason, ``A new method for combining {NLO QCD} with shower {M}onte {C}arlo
  algorithms,'' {\em JHEP} {\bf 11} (2004) 040,
\href{http://www.arXiv.org/abs/hep-ph/0409146}{{\tt hep-ph/0409146}}.

\bibitem{Nason:2006hf}
P.~Nason and G.~Ridolfi, ``A positive-weight next-to-leading-order {M}onte
  {C}arlo for {Z} pair hadroproduction,'' {\em JHEP} {\bf 08} (2006) 077,
\href{http://www.arXiv.org/abs/hep-ph/0606275}{{\tt hep-ph/0606275}}.

\bibitem{Gieseke:2003hm}
S.~Gieseke, A.~Ribon, M.~H. Seymour, P.~Stephens, and B.~Webber, ``Herwig++
  1.0: An event generator for $e^+e^-$ annihilation,'' {\em JHEP} {\bf 02}
  (2004) 005,
\href{http://www.arXiv.org/abs/hep-ph/0311208}{{\tt hep-ph/0311208}}.

\bibitem{Gieseke:2003rz}
S.~Gieseke, P.~Stephens, and B.~Webber, ``New formalism for {QCD} parton
  showers,'' {\em JHEP} {\bf 12} (2003) 045,
\href{http://www.arXiv.org/abs/hep-ph/0310083}{{\tt hep-ph/0310083}}.

\bibitem{Abreu:1996na}
{\bf DELPHI} Collaboration, P.~Abreu {\em et al.}, ``Tuning and test of
  fragmentation models based on identified particles and precision event shape
  data,'' {\em Z. Phys.} {\bf C73} (1996)
11--60.

\bibitem{Gieseke:2006ga}
S.~Gieseke {\em et al.}, ``Herwig++ 2.0 release note,''
\href{http://www.arXiv.org/abs/hep-ph/0609306}{{\tt hep-ph/0609306}}.

\bibitem{Webber:1999ui}
B.~R. Webber, ``Fragmentation and hadronization,''
\href{http://www.arXiv.org/abs/hep-ph/9912292}{{\tt hep-ph/9912292}}.

\bibitem{Acton:1991aa}
{\bf OPAL} Collaboration, P.~D. Acton {\em et al.}, ``A study of charged
  particle multiplicities in hadronic decays of the {Z0},'' {\em Z. Phys.} {\bf
  C53} (1992)
539--554.

\end{thebibliography}\endgroup
\bibliographystyle{utphys}
\end{document}